\documentclass[letter,12pt]{article}
\usepackage{graphicx,amssymb,amsmath}

\def\beq{\begin{equation}}
\def\eeq{\end{equation}}
\def\bea{\begin{eqnarray}}
\def\eea{\end{eqnarray}}
\def\bmat{\begin{pmatrix}}
\def\emat{\end{pmatrix}}
\def\bei{\begin{itemize}}
\def\eei{\end{itemize}}

\def\gev{\, {\rm GeV}}
\def\tev{\, {\rm TeV}}

\newcommand{\gsim}{\lower.7ex\hbox{$\;\stackrel{\textstyle>}{\sim}\;$}}
\newcommand{\lsim}{\lower.7ex\hbox{$\;\stackrel{\textstyle<}{\sim}\;$}}
\def\bq{\begin{quote}}
\def\eq{\end{quote}}

\newcounter{mnotecount}[section]

\begin{document}
\baselineskip=18pt

\begin{titlepage}

\noindent
\begin{flushright}
MCTP-09-39 \\
CERN-PH-TH-2009-154\\
IFT-09-08\\
\end{flushright}
\vspace{1cm}

\begin{center}
  \begin{Large}
    \begin{bf}
    Reheating Temperature and Gauge Mediation Models of Supersymmetry Breaking

        \end{bf}
  \end{Large}
\end{center}

\vspace{0.5cm}
\begin{center}
Marek Olechowski$^{a}$, Stefan Pokorski$^{a}$, Krzysztof Turzy\'nski$^{a}$, James D. Wells$^{b,c}$

\vspace{0.3cm}
\begin{it}
${}^{a}$Institute of Theoretical Physics, Warsaw University, Ho\.za 69, 00-681, Warsaw, Poland\\
\vspace{0.1cm}
${}^{b}$CERN Theory Group (PH-TH), CH-1211 Geneva 23, Switzerland\\
\vspace{0.1cm}
${}^c$MCTP, University of Michigan, Ann Arbor, MI 48109, USA

\end{it}

\vspace{1cm}
\end{center}

\begin{abstract}

For supersymmetric theories with gravitino dark matter, the maximal reheating temperature consistent with  big bang nucleosynthesis bounds arises when the physical gaugino masses are degenerate. We consider the cases of a stau or sneutrino next-to-lightest superpartner, which have relatively less constraint from big bang nucleosynthesis. The resulting parameter space is consistent with leptogenesis requirements, and can be reached in generalized gauge mediation models. Such models  illustrate a class of theories that overcome the well-known tension between big bang nucleosynthesis and leptogenesis.

\end{abstract}

\vspace{3cm}

\begin{flushleft}
\begin{small}
August 2009
\end{small}
\end{flushleft}

\end{titlepage}


\tableofcontents

\section{Introduction}

The question of reconciling supersymmetric dark matter scenarios with the standard evolution of the Universe from very high temperatures was raised long ago~\cite{Ellis:1982yb,Khlopov:1984pf,Ellis:1984eq,Ellis:1984er,Balestra:1986gg,Kawasaki:1994af,Ellis:1995mr,Bolz:2000fu} 
and addressed by many authors since (see, e.g., \cite{Feng:2003zu} for a review).
It is well known that with unstable gravitinos we face the so-called gravitino problem, as for example in generic gravity mediation scenarios of supersymmetry breaking with neutralino lightest supersymmetric partner (LSP). If gravitinos are overproduced, their decay products can destroy  the otherwise successful predictions of Big Bang nucleosynthesis (BBN).
This, in turn, leads to  strong limits on the reheating temperature $T_\mathrm{R}$ (for a recent analysis see, e.g.,~\cite{Roszkowski:2004jd,Cerdeno:2005eu,Pradler:2006hh,Choi:2007rh,Steffen:2008bt,Kawasaki:2008qe}), which may be in conflict with
the higher temperatures required for  thermal leptogenesis (see, e.g., \cite{Davidson:2008bu}). An interesting exception occurs when the gravitino is very
heavy, $m_{3/2}>10\,\mathrm{TeV}$, as in anomaly-mediated scenarios~\cite{Gherghetta:1999sw} 
or mirage mediation models of supersymmetry transmission to the visible
sector~\cite{Choi:2005ge}. Such heavy gravitinos decay before the onset of nucleosynthesis.

In scenarios with stable gravitinos, such as gauge mediated supersymmetry breaking, it is also generically dificult to reach high reheating
temperature while maintaining consistency with the BBN bounds. In this case,  the decay products of the
next-to-lightest supersymmetric particle (NLSP) threaten to alter 
BBN.  Further, if the gravitino is the only constituent of dark matter, which is the case considered in this paper, its cosmological abundance is fixed to be $\Omega_{\tilde G} h^2=0.110\pm0.006$ \cite{Dunkley:2008ie}. Gravitinos can be thermally produced in the post-inflationary universe, with the abundance proportional to the reheating temperature, $\Omega_{\tilde G}^\mathrm{TP}h^2 \propto T_\mathrm{R}$, and also proportional to a factor depending on the precise superpartner spectrum. Requirements on $\Omega_{\tilde G}$ turn into requirements on $T_\mathrm{R}$ for a given superpartner spectrum.

The BBN constraints have been extensively studied both in
a model-independent way and for specific choices of the NLSP. By a model-independent
approach, we mean limits on $Y_\mathrm{NLSP}$, which is defined to be the number density normalized to entropy density $n_\mathrm{NLSP}/s$, as a function of the NLSP lifetime 
$\tau_\mathrm{NLSP}$, the NLSP mass $m_\mathrm{NLSP}$ and the branching ratios of its
electromagnetic and hadronic decays (see, e.g., \cite{Kawasaki:2004qu}). For specific choices of the NLSP, this analysis can be done
in a more precise way, since the decay modes of the NLSP can be studied in detail. Most of the interesting NLSP candidates have already been discussed in the literature:
neutralinos, staus, sneutrinos, stops and gluinos (see, e.g., \cite{Roszkowski:2004jd,Cerdeno:2005eu,Kawasaki:2008qe,Ratz:2008qh,Pradler:2008qc,Ellis:2008as,Covi:2007xj,Berger:2008ti,Bailly:2009pe}).

Some general points are worthy of introductory note.
First, the NLSP lifetime depends inversely on the ratio of the  NLSP to gravitino masses. Since we wish to have a fast decay lifetime so that decay products can thermalise before BBN finishes its work,
one gets a lower bound for this ratio $m_{\rm NLSP}/m_{3/2}$ and, hence, also for the rest of the spectrum.
Precise numbers depend on details of the spectrum. Also, 
more recently,  it has been pointed out~\cite{Pospelov:2006sc} that a very stringent bound exists for a charged NLSP from potential overproduction of ${}^6\mathrm{Li}$~\cite{Dimopoulos:1989hk}.   And finally, previous studies indicate that it is  difficult
to reconcile gravitino dark matter with a reheating temperature high enough for successful thermal leptogenesis (for a recent  discussion see~\cite{Choi:2007rh,Steffen:2008bt}). 

Our goals are to put details on all these points and more, and show that it is possible for constraints to be met in the context of a high reheat temperature needed for leptogenesis. 
We first systematically search for the patterns of supersymmetric spectra with gravitino LSP that maximize the reheating temperature consistently with the required gravitino relic abundance and the BBN bounds. Since the stau NLSP ($\tilde\tau$) and sneutrino NLSP ($\tilde\nu$) have the
smallest hadronic branching ratios, the BBN bounds are the weakest for these choices of light NLSP and we restrict our subsequent analysis to these two cases.  
One of the main features of the obtained spectra is that compatibility with all conditions and maximizing the reheat temperature pressures us away from universal gaugino masses at the high scale. In the second part of the paper we  apply the techniques of general gauge mediation to see whether such spectra can indeed be obtained within a model of gauge mediation of supersymmetry breaking. 

\section{Maximal reheating temperature with stau or sneutrino as NLSP}

Gravitinos constituting the dark matter can be produced both thermally after inflation and nonthermally from NLSP decays. We shall focus first on the thermal
component which can be written as \cite{Bolz:2000fu,Pradler:2006qh}:
\begin{equation}
\label{c1}
\Omega_{\tilde G}^\mathrm{TP}h^2 = \left(\frac{m_{3/2}}{1\,\mathrm{GeV}}\right)\left(\frac{T_\mathrm{R}}{10^{10}\,\mathrm{GeV}}\right)\sum_r y'_r g_r^2(T_\mathrm{R})(1+\delta_r)\left( 1+\frac{M^2_r(T_\mathrm{R})}{3m^2_{3/2}}\right)\ln\left(\frac{k_r}{g_r(T_\mathrm{R})}\right)\, .
\end{equation}
The sum runs over the Standard Model gauge groups and $r=1,2,3$
corresponds to $U(1)_Y$, $SU(2)_L$, $SU(3)_C$, respectively.
The values of the coefficients $y'_r$  and $k_r$ can be extracted from~\cite{Pradler:2006qh}, and the
coefficients $\delta_r$ parametrize the corrections to this result coming from novel gravitino
production channels opening up in the presence of finite temperature corrections \cite{Rychkov:2007uq}. Numerically, $\delta_r$ are  $\sim$0.1, 0.2, 0.4 for $r=1,2,3$ and $T_\mathrm{R}$ between $10^7$ and $10^9\,\mathrm{GeV}$. 

We are interested in maximizing  the reheating temperature in eq.\ (\ref{c1}) for 
$\Omega_{\tilde G}^{TP} h^2=0.11$. We shall come back shortly to the role of nonthermal production, but suffice it to say now that it plays a smaller role. We require consistency  with all available  bounds, particularly the BBN bound. It is natural and convenient to study the maximal possible reheating temperature as a function of the physical NLSP mass. This mass acts as the kinematical upper bound on the gravitino mass and as the lower bound for the rest of the physical masses of the MSSM spectrum. Furthermore, as we remind the reader later on,  the BBN bounds can be expressed as  bounds for the gravitino mass as a function of the NLSP mass. It is therefore convenient to rewrite the relation (\ref{c1})  in terms of the physical masses, with the NLSP mass introduced as the  reference scale for the other masses:
\begin{equation} 
\label{c1a}
\Omega_{\tilde G}^\mathrm{TP}h^2 =\left(\frac{T_\mathrm{R}}{10^{9}\,\mathrm{GeV}}\right)\left(\frac{m_\mathrm{NLSP}}{300\,\mathrm{GeV}}\right)
\left[7.4\times 10^{-6}\frac{\frac{m_{3/2}}{1\,\mathrm{GeV}}}{\frac{m_\mathrm{NLSP}}{300\,\mathrm{GeV}}}+\frac{\frac{m_\mathrm{NLSP}}{300\,\mathrm{GeV}}}
{\frac{m_{3/2}}{1\,\mathrm{GeV}}}
\sum_r \gamma_r\left(\frac{M_r}{m_\mathrm{NLSP}}\right)^2 \right] \, ,
\end{equation}
where $M_r$ denote physical gaugino masses and the coefficients $\gamma_r$ depend on the ratios of the gauge couplings at the reheating scale and 
the scale of the physical gaugino masses.  The values of $\gamma_r$ can be evaluated
for $T_\mathrm{R}=10^9\,(10^7)\,\mathrm{GeV}$ as $\gamma_3=0.48-0.56\,(0.62-0.74)$, $\gamma_2=0.57\,(0.54)$, $\gamma_1=0.22\,(0.17)$, where the range for $\gamma_3$ corresponds to
the gluino masses ranging from  200 to 900 GeV. We have used here the 1-loop RGE for the gaugino masses.

It is clear  that the first condition for maximizing the reheating temperature for fixed $\Omega_{\tilde G}^\mathrm{TP}h^2$  and fixed NLSP mass is that  the sum in eq.~(\ref{c1a}) has its minimal value, i.e.\ that the  gaugino masses are completely degenerate with the NLSP mass. Degenerate physical gaugino masses mean, of course, that they cannot be degenerate at a high scale. This also means that the gluon and gluino thermal scattering is no longer the dominant source of gravitino thermal production, as would be  the case if  gaugino masses were universal at the high scale. 
This information is directly encoded
in the values of  the coefficients $\gamma_r$, which in turn, has serious implications for the constraints of the maximal reheating
temperature consistent with the BBN bounds. This reasoning has been previously applied for
constraining the gluino mass by successful leptogenesis \cite{Fujii:2003nr}.

Secondly, the maximal reheating temperature for a fixed $m_\mathrm{NLSP}$  is obtained for a gravitino mass that minimizes the square bracket in eq.~(\ref{c1a}).  
For a given type and mass of the NLSP, this value of $m_{3/2}$ may imply an NLSP lifetime
for which its relic thermal abundance
(before it decays) violates the BBN bounds.
We must, therefore, discuss when this happens and find the maximal reheating temperature
corresponding to the gravitino mass (or, equivalently, the NLSP lifetime) consistent with the
BBN bounds.

Very generally, the BBN bounds require that unstable relics previously present in the Universe, decay
with lifetimes smaller than 100 s, unless the abundance of these particles is very small or only a tiny
fraction of these particles decay with energetic hadrons in the final state~\cite{Kawasaki:2004qu}. 
Among the MSSM particles,
the latter condition is satisfied by the lightest sneutrino or the lightest stau, hence its parameter space allows for $\tau_{\rm NLSP}>100\, s$.

\begin{figure}
\begin{center}
\includegraphics*[height=7.8cm]{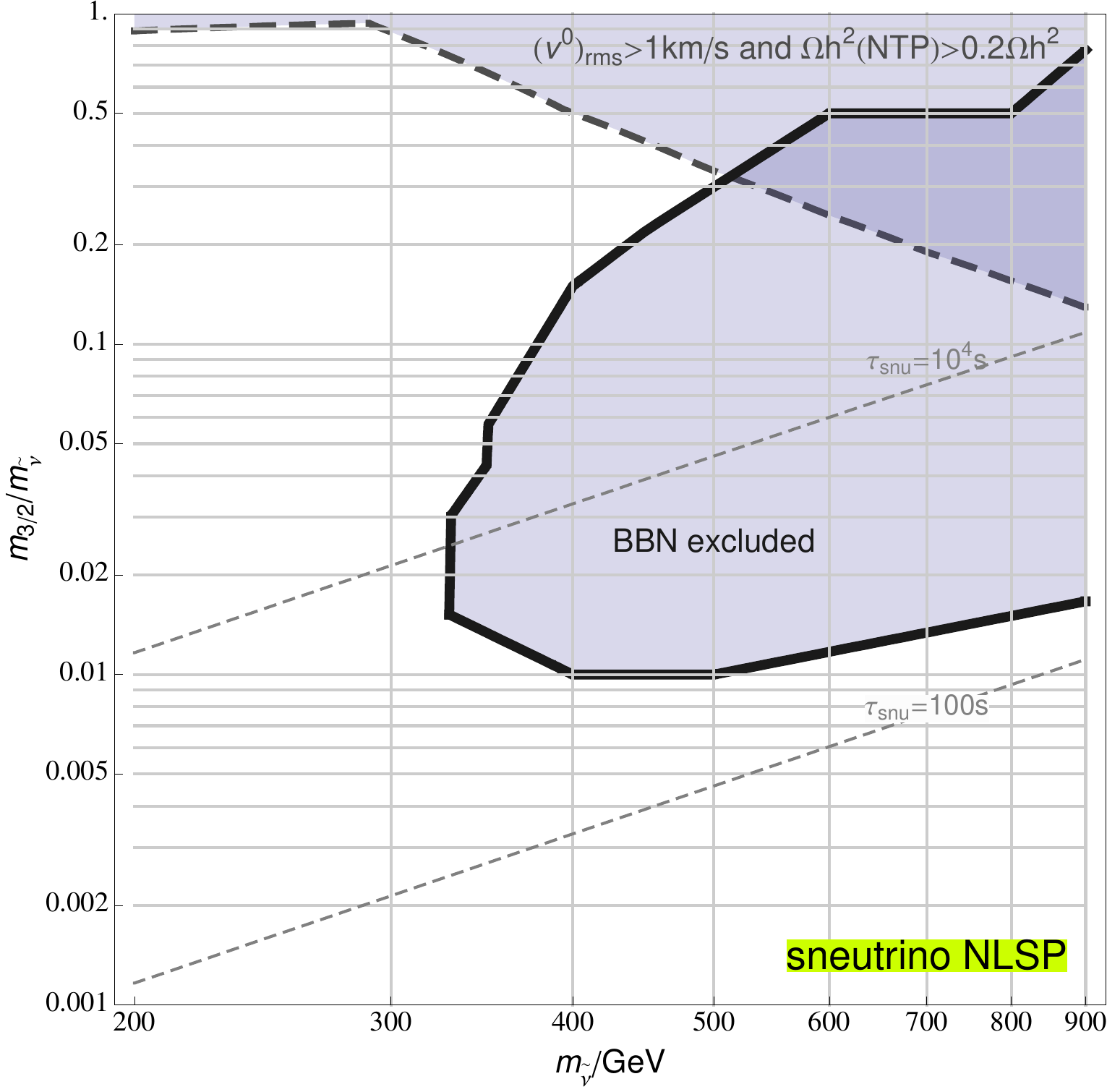}
\hspace{0.5cm}
\includegraphics*[height=7.8cm]{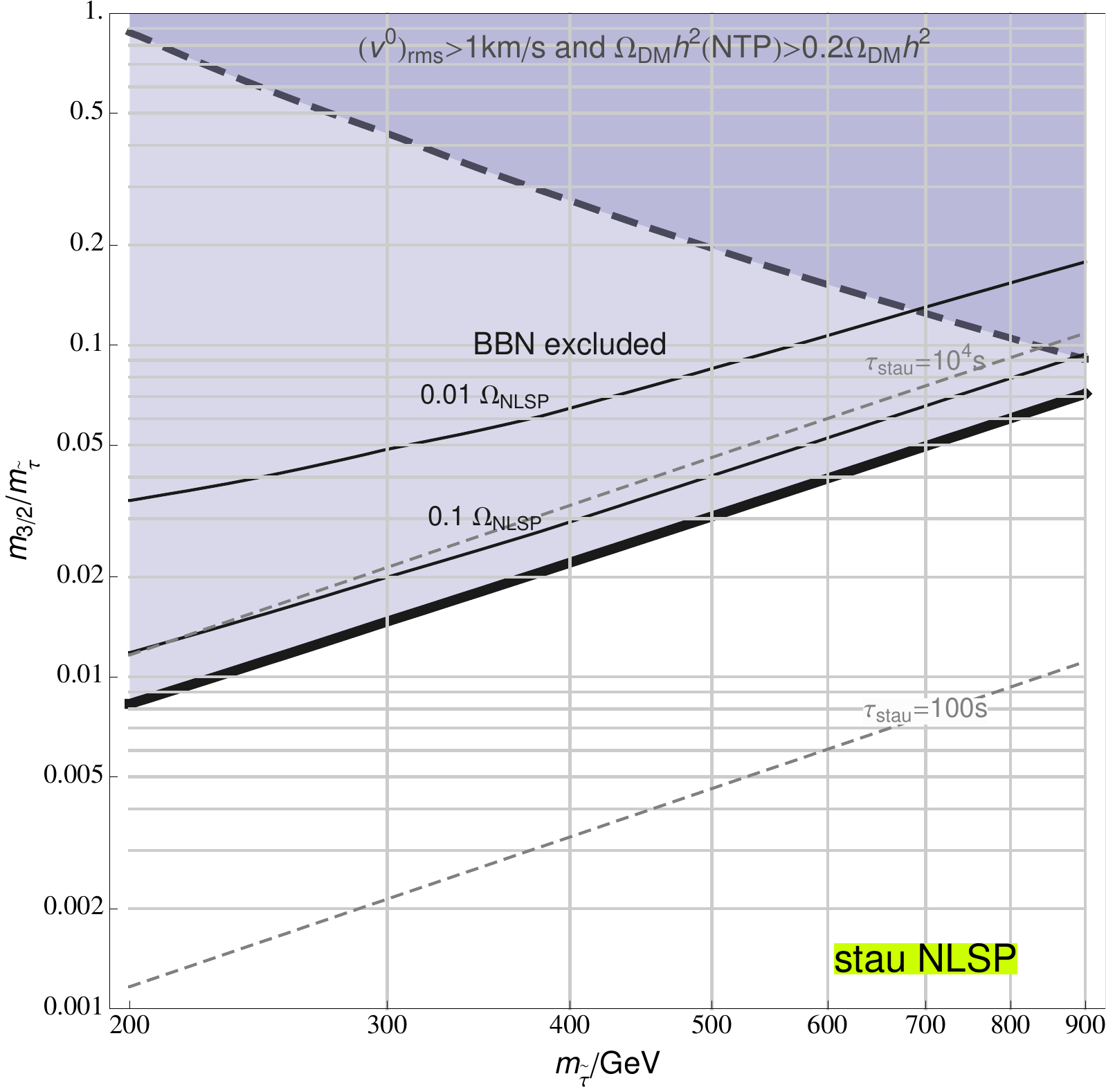}
\end{center}
\caption{\em {\bf Left panel:} for sneutrino NLSP, we draw the BBN bounds adapted from~\cite{Kawasaki:2008qe} (shaded exclusion region inside the solid line), and  the large scale structure bounds in the mixed dark matter scenario adapted from~\cite{Jedamzik:2005sx} (shaded exclusion region above the dashed line). We also show contours of 
 constant NLSP lifetime $\tau =100\, s$ and $10^4 s$ (short-dashed). {\bf Right panel:} Same as left panel except the NLSP is stau and the BBN bounds are calculated with the use of the exclusion plots of~\cite{Pradler:2007is}.  $0.01\Omega_{\rm NLSP}$ and $0.1\Omega_{\rm NLSP}$ lines indicate what the bound would be if the NLSP would-be relic abundance were lowered, e.g.\ by co-annihilation, by factors of $0.01$ and $0.1$ respectively. In these figures, $m_{\rm snu}\equiv m_{\tilde\nu}$ is the sneutrino NLSP mass and $m_{\rm stau}\equiv m_{\tilde\tau}$ is the stau NLSP mass. \label{f1}}
\end{figure}

The cosmological constraints on late neutrino injection have been worked out~\cite{Kanzaki:2007pd} and later  updated and specified to the sneutrino NLSP case~\cite{Kawasaki:2008qe,Jedamzik:2007qk}. It has been found that sneutrinos with masses smaller than about 330 GeV
evade the BBN constraints. Such light sneutrinos are mostly constrained by the requirement that large scale structure formation is not too much affected by free-streaming gravitinos produced in the sneutrino decays~\cite{Jedamzik:2005sx}. Numerically, this constraint is very similar to that resulting from  the scenario's founding proposition that the gravitino is lighter than the sneutrino,  $m_{3/2}<m_{\tilde\nu}$. As we shall see shortly, the lighter the sneutrino the higher is the reachable  reheating temperature and therefore we shall restrict our considerations to a sneutrino in the mass range between 200 and 330 GeV.  There is some uncertainty in the literature regarding the constrained region from BBN for sneutrinos (see, e.g., ref.~\cite{Covi:2007xj}). For example, the excluded thumb region in Fig.~\ref{f1}, which we shall discuss shortly, may be uncertain to the left or right by $\sim 50\gev$. This level of uncertainty does not upset our analysis, as the region of light NLSP and maximal reheating temperature is above the thumb region for the lower sneutrino mass values.

For the sake of completness, we further emphasize how for heavier sneutrino NLSP the BBN constraints are especially restrictive. For $m_{\tilde\nu}$ between ~300  and 500 GeV  the requirement that ${}^6\mathrm{Li}$ and D are not
overproduced excludes gravitino masses between a few and a few tens of GeV \cite{Kawasaki:2008qe}.
The assumption that the hot component of the gravitino dark matter originating from the NLSP decays
makes up at most 20\% of the total dark matter energy density \cite{Jedamzik:2005sx} forbids $m_{3/2}$
too close to $m_{\tilde\nu}$. Finally, for  still larger values of $m_{\tilde\nu}$, between 500 and 900 GeV, only solutions with light gravitino, with mass between a few and 10 GeV, remain. Since only light sneutrinos are of interest for
us, there is no need to pursue a more quantitative study of how the BBN bounds depend on the mass spectrum of the MSSM. 

For stau NLSP with masses less than 1 TeV, the main BBN constraint is that of a correct primordial ${}^6\mathrm{Li}$ abundance coming from catalyzed ${}^6\mathrm{Li}$
production \cite{Dimopoulos:1989hk,Pospelov:2006sc}. We conservatively take it as ${}^6\mathrm{Li}/\mathrm{H}<6\times 10^{-11}$ \cite{Steffen:2008bt}. This provides a constraint on the relic stau abundance as a function of its lifetime~\cite{Pradler:2007is}. Staus that decay faster than $5\times 10^3\, {\rm s}$ are generally safe and independent of relic abundance~\cite{Pospelov:2006sc,Pradler:2008qc} (but see \cite{Bailly:2009pe} for exceptions). We calculate the relic stau abundance with all other superpartners decoupled, and we use this result to obtain the  upper bound on the  gravitino mass as a function of the stau mass, while remaining consistent with the BBN bound.  We stress that these bounds are rather robust: had coannihilations reduced the relic stau abundance by a factor of 10 (or 100, as is possible with an extreme gluino/stau degeneracy), the resulting bound on the gravitino
mass  would increase by a factor of 1.3-1.5 (3-4)\footnote{This also shows that the bounds on the
allowed $m_{3/2}$ and $m_{\tilde\tau}$ mass ranges would change little if we used a somewhat more (less) conservative ${}^6\mathrm{Li}/\mathrm{H}$ bound.}.
The  BBN bounds for sneutrino and stau NLSP discussed here are pictorially summarized by Fig.~\ref{f1} in the plane $m_{3/2}/m_\mathrm{NLSP}$ vs.\ $m_\mathrm{NLSP}$.

For light sneutrinos considered in this paper, the only constraint is
$m_{3/2}<m_\mathrm{NLSP}$, and both terms in the square bracket in  eq.~(\ref{c1a}) are important in its minimization with respect
to the gravitino mass. The obtained values of the gravitino mass are slightly smaller than the NLSP mass, with their ratio being almost independent of the NLSP mass.  The latter property suggests, from inspection of eq.\ (\ref{c1a}), that the maximal
reheating temperature will decrease with increasing NLSP mass.

For stau NLSP, as seen from Fig.\ \ref{f1}, the BBN bound puts an upper bound on the gravitino mass much below the stau mass.
The first term in eq.\ (\ref{c1a}) is therefore negligible and the dependence of the reheating temperature on the NLSP mass
is governed by the behaviour of the ratio of the maximal gravitino mass consistent with BBN to the NLSP mass multiplied by a linear function of the latter. As seen in Fig.\ \ref{f1}, this ratio  increases faster than the NLSP mass itself. Therefore, the maximal reheating temperature is expected to rise with the stau mass, in contrast to what we found for the sneutrino LSP.  Our plot begins at $m_{\tilde\tau}=200\gev$ since, as we shall see, experimental limits require it after taking into account correlations with the other superpartner masses. 

We are now in a position to calculate the maximal reheating temperature for the sneutrino and stau NLSP. Using eq.~(\ref{c1a}) and including the gravitinos produced nonthermally in the NLSP decays, we obtain
\begin{equation}
\Omega_{\tilde G} h^2 = \Omega_{\tilde G}^\mathrm{TP} h^2+\frac{m_{3/2}}{m_\mathrm{NLSP}} \Omega_\mathrm{NLSP}h^2 \, .
\label{eq:full omega}
\end{equation}
Here, $\Omega_\mathrm{NLSP}$ represents the thermal relic abundance that the NLSP would have had, had it not decayed into gravitinos and Standard Model particles. We calculate the value of this
parameter with the {\tt micrOMEGAs.2.2} code \cite{Belanger:2006is,Belanger:2008sj}, assuming
that only the NLSP is light  while all the other
supersymmetric particles have masses of 2 TeV. This choice is meant to eliminate nongeneric
coannihilations which may be inherent features of particular scenarios of supersymmetry breaking.
It turns out that in the parameter range that maximizes the reheating temperature nonthermal gravitino production is always well below 20\% of the total, and our qualitative discussion of the thermal  production explains the full results.

\begin{figure}
\begin{center}
\includegraphics*[height=7.0cm]{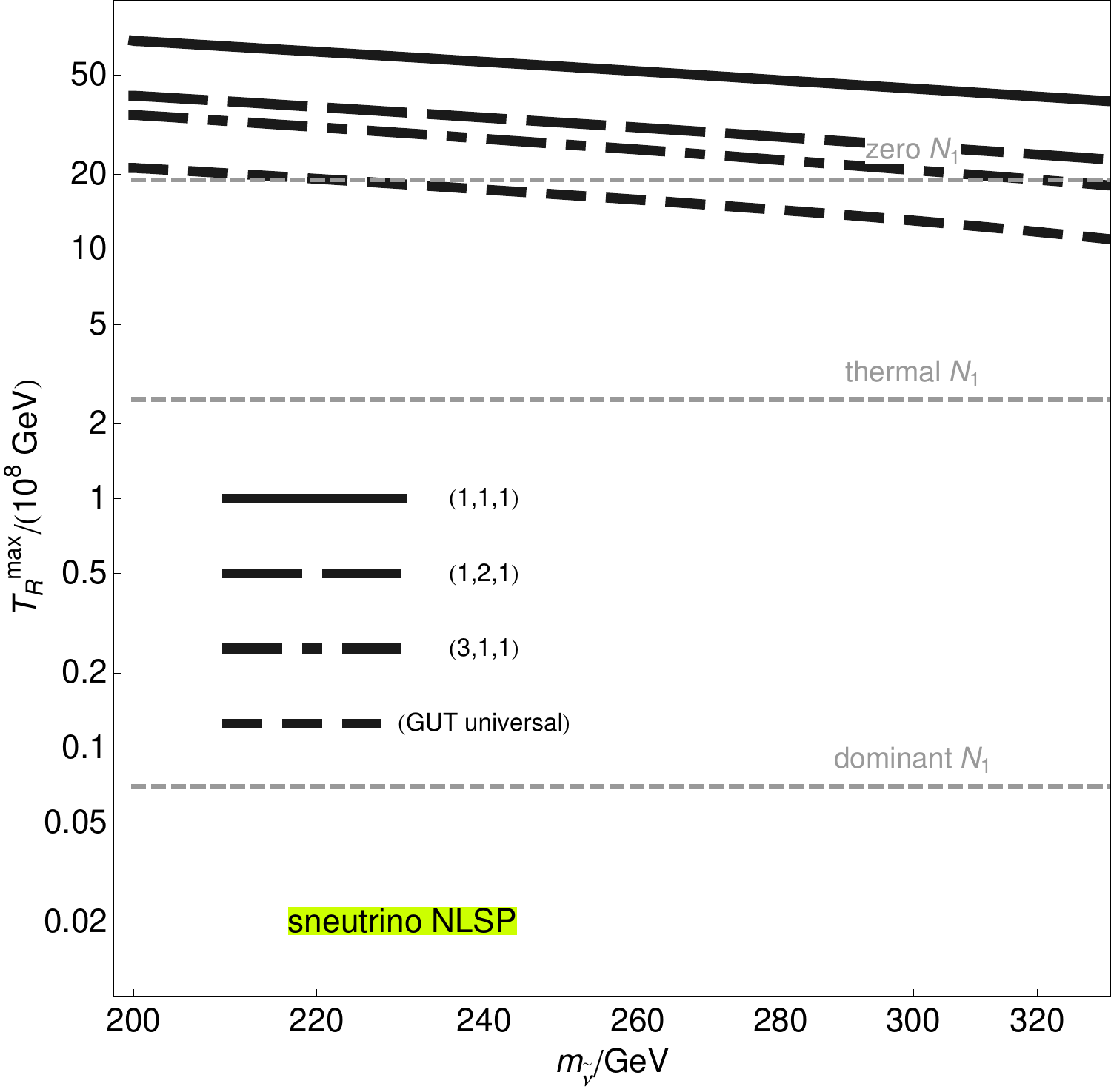}
\hspace{0.5cm}
\includegraphics*[height=7.0cm]{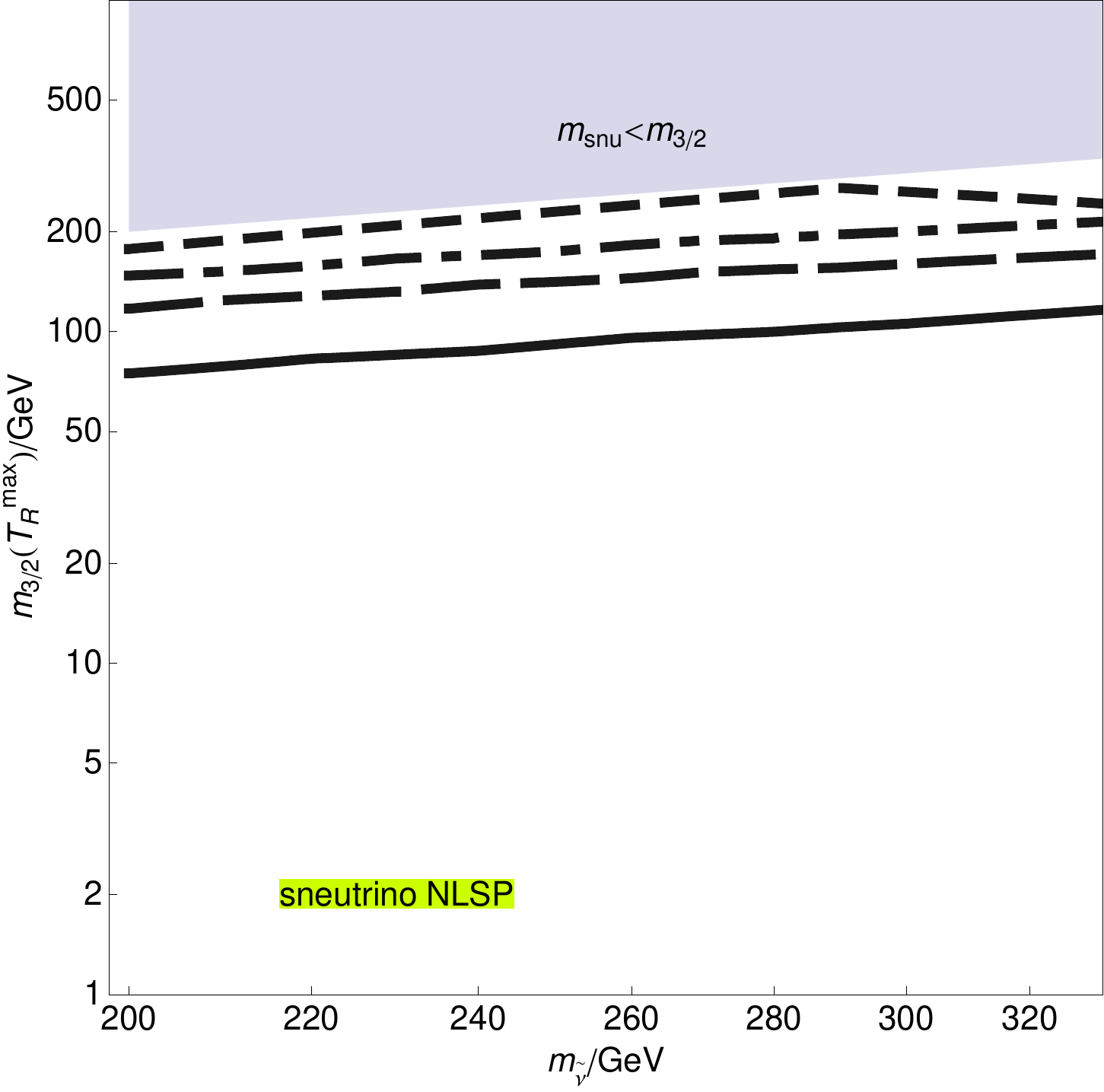}
\end{center}
\caption{\em Sneutrino NLSP: the maximal reheating temperature (left) and the gravitino mass corresponding to the maximal reheating temperature (right) for four mass patterns of the gauginos at the low-scale $(M_3/m_\mathrm{NLSP},M_2/m_\mathrm{NLSP},M_1/m_\mathrm{NLSP})=(1,1,1),\,(1,2,1),\,(3,1,1),$ and $({\rm GUT\, universal})$. They correspond, respectively, to  solid,  dash-dotted, long-dashed, and  short-dashed lines. `Dominant', `zero' and `thermal' $N_1$ lines correspond to lower limits of $T_\mathrm{R}$ needed for leptogenesis given various scenarios discussed in the text. In this figure, $m_{\rm snu}\equiv m_{\tilde\nu}$ is the sneutrino NLSP mass. \label{bbnsnu}}
\end{figure}

\begin{figure}[t]
\begin{center}
\includegraphics*[height=7.8cm]{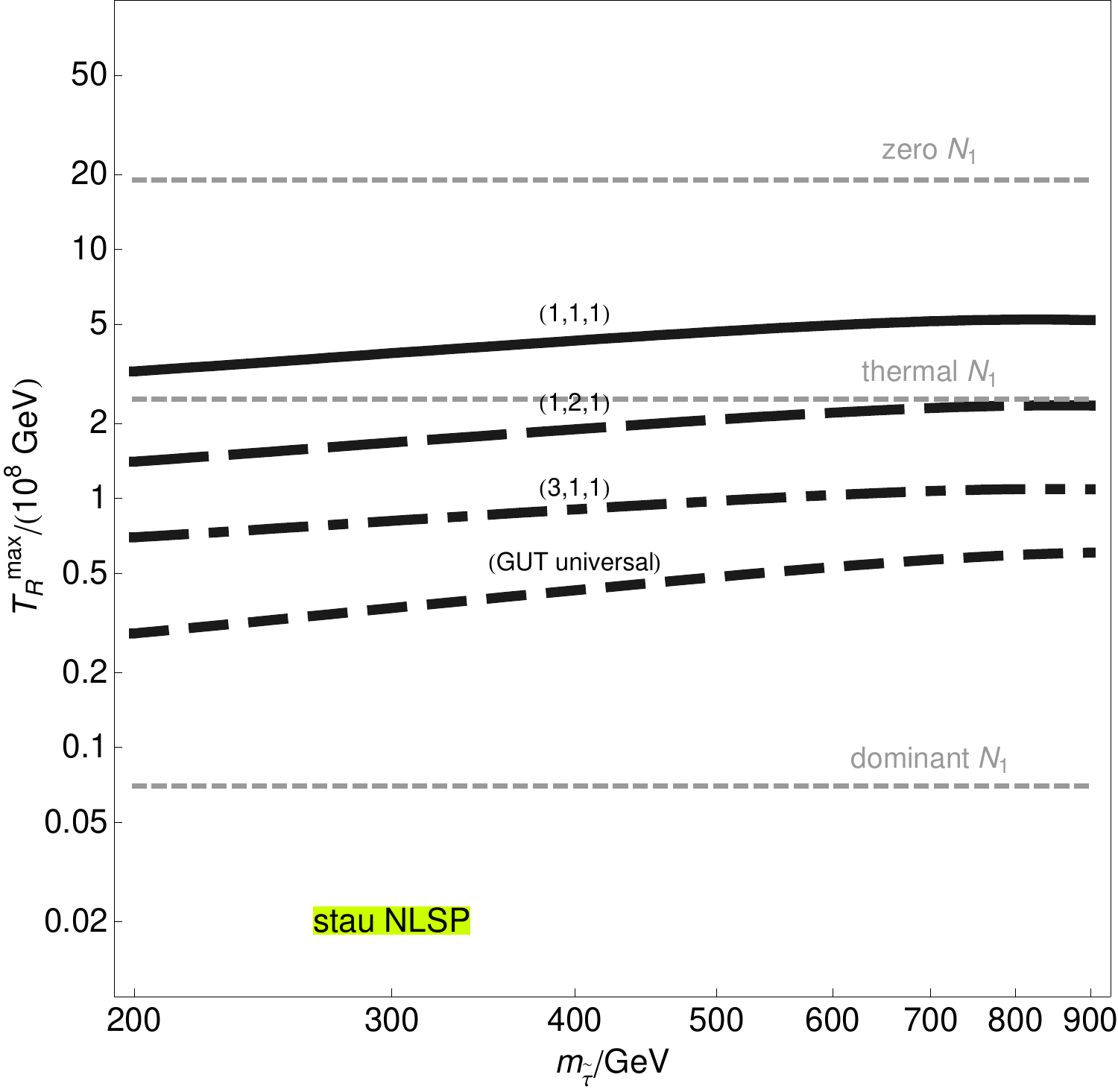}
\end{center}
\caption{\em Stau NLSP: the maximal reheating temperature for four mass patterns of the gauginos at the low-scale $(M_3/m_\mathrm{NLSP},M_2/m_\mathrm{NLSP},M_1/m_\mathrm{NLSP})=(1,1,1),\,(1,2,1),\,(3,1,1),$ and $({\rm GUT\, universal})$. They correspond, respectively, to  solid, dash-dotted,  long-dashed, and short-dashed lines. `Dominant', `zero' and `thermal' $N_1$ lines correspond to lower limits of $T_\mathrm{R}$ needed for leptogenesis given various scenarios discussed in the text. 
In this figure, $m_{\rm stau}\equiv m_{\tilde\tau}$ is the stau NLSP mass. \label{bbnstau}}
\end{figure}

 The maximum $T_\mathrm{R}$ is then obtained by requiring that $\Omega_{\tilde G} h^2\leq 0.11$ from eq.~(\ref{eq:full omega}), subject to the constraining relationship between the gravitino and NLSP masses.  For the case of sneutrino NLSP with mass less than $330\gev$, the only constraint is $m_{3/2}<m_{\tilde\nu}$. For the case of stau NLSP 
this implies the requirement that $m_{3/2}\ll m_{\tilde\tau}$. Therefore, we expect the maximum reheat temperature for sneutrinos to be much higher than for stau since the ratio $m_{\tilde\nu}/m_{3/2}$, which is so important in eq.~(\ref{c1a}), can be much lower than $m_{\tilde\tau}/m_{3/2}$, enabling a compensating $T_\mathrm{R}$ to be much higher. The results for the maximal reheating temperature with sneutrino (stau) NLSP and for the gravitino mass corresponding to this temperature are shown in Fig.~\ref{bbnsnu}\, (Fig.~\ref{bbnstau}) for four characteristic patterns of gaugino masses. We learn from Fig.~\ref{bbnsnu} that for the sneutrino NLSP the higher the reheat temperature contour, such as the solid line, the lower the gravitino mass, and it is not expected that the gravitino mass be nearly degenerate with the NLSP mass. 

In Fig.~\ref{bbnsnu} and~\ref{bbnstau}, we also plot for reference  minimal reheating temperatures needed for leptogenesis given various simple assumptions about the details of the reheating mechanism. These reference values are taken from ref.~\cite{Giudice:2003jh,Antusch:2006gy}. The line `zero $N_1$' corresponds to the result $T_\mathrm{R}>1.9\times 10^9\,\mathrm{GeV}$. This $T_\mathrm{R}$ value is obtained by assuming the initial condition after inflation that there are zero $N_1$ particles, or in other words inflaton decay yields every  kinematically accessible particle except the heavy (s)neutrinos. 
The line `thermal $N_1$'  corresponds to the bound $T_\mathrm{R}>2.5\times 10^8\,\mathrm{GeV}$. This assumes the initial condition  that all species, including the lightest right-handed (s)neutrinos $N_1$ but not species heavier than that, fill the universe and are in thermal equilibrium after inflation.
`Dominant $N_1$' implies the lower reheating bound of 
$T_\mathrm{R}>7\times 10^6\,\mathrm{GeV}$. `Dominant $N_1$'  assumes that only the lightest right-handed (s)neutrinos $N_1$ have the initial condition of thermal equilibrium abundance after inflation, while all other particles are initially absent.

There are effects that can make the bounds weaker and effects that make the bound stronger when some of our assumptions are altered.
For example, we assume a very hierarchical mass spectrum among the right-handed neutrinos.  Bounds can be significantly relaxed for a mildly \cite{Hambye:2003rt,Raidal:2004vt}, very~\cite{Flanz:1994yx} or completely~\cite{Turzynski:2004xy} degenerate mass spectrum.  It must also be acknowledged that 
maximally efficient leptogenesis corresponds to a very particular corner of the parameter space of the seesaw mechanism; hence, in a realistic model the lower bounds on $T_\mathrm{R}$ given above may be generally higher.  Yet another example, the `zero $N_1$'
bound assumes no direct production of the right-handed neutrinos during reheating, which may not be a realistic option since  couplings are likely to be generic between the inflaton and the right-handed neutrinos, which are all Standard Model singlets. Therefore, leptogenesis bounds quoted here should be taken as indicative predictions of particularly simple scenarios rather than true cosmological constraints. On the other hand, given that the range of `leptogenesis-friendly' reheating temperatures partially overlaps with the range of $T_\mathrm{R}$ giving a correct dark matter abundance consistent with the BBN, it is interesting to study scenarios in which the reheating temperature is maximized, as they are potentially least constrained by cosmology.

There are two more well-known and universal constraints on the superpartner spectrum that must be taken into account. First, the lower experimental limit on the Higgs boson mass implies a lower bound on the geometrical average of the  stop masses. This is because raising the Higgs mass is accomplished to leading order by a loop factor proportional to the logarithm of the stop masses.  For small values of the $A$-terms, as in gauge mediation models,  the lower bound is about 1 TeV. This bound gains special significance for model building in the presence of light gluinos, as we will see below. A challenging issue we will confront is the difficulty in obtaining small degenerate gaugino masses and stau or sneutrino NLSP while simultaneously producing squark masses heavy enough to lift the Higgs boson above the current experimental limit. The second constraint, which will be addressed below,  is requiring that the higgsino mass be heavier than the NLSP mass. The higgsino mass is governed by the $\mu$ parameter whose value must be consistent with a proper electroweak symmetry breaking potential.

In summary, we see that  with stau as the NLSP, the maximal reheating temperature ${\rm few}\times 10^8\gev$ can be reached when its mass is greater than $200\gev$ and almost degenerate  with all the  gaugino  masses, and the gravitino is in  the mass range $\mathcal{O}(1-10)\,\mathrm{GeV}$. For sneutrino NLSP,
the reachable reheating temperature $\gsim 10^9\gev$  is certainly in a range interesting for thermal leptogenesis. Common to both cases of stau or sneutrino NLSP, the stop masses are around $1\tev$ in order to satisfy current limits on the Higgs boson mass, and the rest of the spectrum is not very constrained from the purely low-energy point of view. In particular, the obvious requirement that  $Y_\mathrm{NLSP}$ is as small as possible is of lesser importance, as discussed earlier.

\section{UV initial conditions in general gauge mediation models\label{UV Conditions}} 

Our next question is if
such spectra can be obtained in gauge mediation models
\cite{GMfirst01,GMfirst02,GMfirst03,GMfirst04,GMfirst05,GMfirst06,GMfirst07,GMfirst1,GMfirst2,GMfirst3,Dudas:2008eq} (see also \cite{GMrev} for a review). 
In supersymmetric models with gauge mediation of supersymmetry breaking, the
gravitino is a well motivated dark matter candidate. It is automatically the
lightest supersymmetric particle (LSP), stable if $R$-parity is conserved.
Indeed, dominance of gauge mediation over gravity mediation implies a gravitino LSP,
as can be readily seen, e.g. in the simplest gauge mediation model.
With supersymmetry broken by a spurion $X=v+F\theta^2$ coupled to a pair
$\mathbf{5}+\mathbf{\overline{5}}$ of messengers, the gauge mediation contribution to the soft scalar 
and gaugino masses is given by the scale $M_\mathrm{susy}=\frac{\alpha}{4\pi}\frac{F}{v}$,
and the gravity mediation contribution  to the 
gaugino and scalar masses  is of the order of the gravitino mass $m_{3/2}$.

We seek a clear dominance of gauge mediation over gravity mediation sufficient to suppress dangerous FCNC transitions with random flavour changing insertions of the order of the gravitino mass~\cite{Nir,Hiller:2008sv,Cerdeno:2009ns}. This requirement implies
$M_\mathrm{susy}/m_{3/2}> 10^{2}$. Furthermore, since $m_{3/2}=\frac{F}{\sqrt{3}M_P}$, and the gravitinos must be in the $\mathcal{O}(1-10)\,\mathrm{GeV}$ range, it is clear that  for soft masses in the TeV mass range, the messenger mass scale must be high, of the order of the GUT scale\footnote{In particular, this suppression of the FCNC disfavors the solution with $m_{3/2}$ too close to the sneutrino NLSP mass.}. For definiteness in examples below, we fix this scale to be either  $10^{14}$ GeV or $10^{15}$ GeV. These choices are conservative, in the sense that they allow for heavy gravitinos and high reheating temperatures. Interestingly,  values $v\ll \alpha M_P$ for the spurion vev are reachable in generic dynamical O'Raifeartaigh-type models of supersymmetry breaking coupled to gravity \cite{Lalak:2008bc}.

Our low-energy constraints for reaching maximal $T_\mathrm{R}$   can  be translated  into conditions for the soft supersymmetry breaking masses $\tilde m_i$ at the messenger mass scale by using the RG evolution.  We run the RG equations downwards in energy, ensuring proper electroweak symmetry breaking, and then study the correlations among other low-energy mass states. 
The choice of degenerate physical gaugino masses at the low scale, which maximizes the reheat temperature, implies  that gaugino masses at the messenger scale do not satisfy the ``universal" initial conditions as in the minimal gauge
mediation models. Rather, their ratios at the high scale are approximately inversely proportional to the squares of the gauge couplings at $M_{\rm susy}$.  To realize this boundary condition one has to study generalized gauge mediation models~\cite{Meade:2008wd,Carpenter:2008wi}. It is therefore convenient to introduce already at this stage the most general parametrization of soft  masses at the messenger mass scale  in general gauge mediation models and to continue
with the RG evolution using this parametrization. We shall later point out the qualitative conclusions, independent of the  chosen high scale parametrization of the soft terms.

In general gauge mediation (GGM), the soft masses at the gauge mediation scale are given by
\begin{eqnarray}
\label{e1}
\tilde M_r &=& \frac{g_r^2}{16\pi^2} \Lambda_r \\
\label{e2}
\tilde m_s^2 &=& 2 \sum_{r=1}^3 \left(\frac{g_r^2}{16\pi^2}\right)^2C_r^{(s)} \tilde\Lambda_r^2 \, ,
\end{eqnarray}
where $r=1,2,3$ corresponds to gauge groups $U(1)_Y$, $SU(2)_L$ and $SU(3)_C$, respectively.
For easier reference, the values of the Casimir invariants are shown in Table \ref{tcas}.
One can also use another description of the GGM models, defining $\kappa_r\equiv \tilde\Lambda_r^2/\Lambda_r^2$ .
The renormalization group equations can be solved semi-analytically for small and moderate 
$\tan\beta$ by means of the bottom-up method of Ref.\ \cite{bu}, and we can then express the scalar masses at the low scale in terms of  large scale values of gaugino masses and $\kappa$'s. 
We apply the RG evolution setting $Q=10^{15}\,\mathrm{GeV}$ or $Q=10^{14}\,\mathrm{GeV}$ as the messenger mass scale. We take the gauge couplings at $M_Z$ as $g_1^2=0.21$, $g_2^2=0.42$, $g_3^2=1.48$, the running top quark mass as $166\,\mathrm{GeV}$, we admit $-10\%$ of supersymmetric threshold corrections~\cite{Pierce:1996zz} to $g_3^2$ and we solve the RGE's between $Q$ and $M_\mathrm{susy}=1000\,\mathrm{GeV}$. 
Since the RG evolution of the gaugino masses is very simple, it turns out to be convenient to have ``hybrid" expressions for low scale values of the soft scalar masses in terms  of the low scale gaugino masses and high scale 
parameters $\kappa$. They are collected in  Appendix A. 

\begin{table}
\begin{center}
\begin{tabular}{c|ccccc}
\hline
\hline
 & $Q$ & $U$ & $D$ & $L,H_u,H_d$ & $E$ \\
\hline
$SU(3)_C$ & 4/3 & 4/3 & 4/3 & 0 & 0 \\
$SU(2)_L$ & 3/4 & 0 & 0 & 3/4 & 0 \\
$U(1)_Y$ & 1/60 & 4/15 & 1/15 & 3/20 & 3/5 \\
\hline
\hline
\end{tabular}
\end{center}
\caption{\em Casimir invariants for the MSSM fields. \label{tcas}}
\end{table}

We can now use the solutions of the RGEs from Appendix A to discuss qualitatively the pattern of supersymmetry breaking consistent with a high reheating temperature.
If stau is to be the NLSP,  it is the lighter stau state, obtained after diagonalization of the mass matrix
with left and right entries. The left entry is also the sneutrino soft mass. From the stau mass matrix, one can see that for
$m_E^2<m_L^2$ the lighter stau is indeed lighter than the sneutrino and we discuss this case first. 

To a good approximation
we can identify the lighter stau with the right stau.
Imposing the bound that the right stau is lighter than the bino, $m_E^2<M_1^2$,  where (see  Appendix A)
\begin{equation}
\label{c5}
m^2_E=(0.56 + 4.9\kappa_1) M^2_1 \,,
\end{equation}
we find
\begin{equation}
\kappa_1 < 0.089
\end{equation}
or, equivalently,  $\tilde m^2_E<0.1 \tilde M^2_1$ at the high scale.  Since in gauge mediation models $\tilde m^2_E>0$,
we also have $M_1^2>m^2_E>0.6 M_1^2$. Thus, the low energy masses of stau and bino are almost degenerate,
as a result of the RG evolution itself, in agreement with the required degeneracy discussed in sec.2.
This conclusion does not depend on the details of the parametrization, eqs.\ (\ref{e1}) and (\ref{e2}).
With $Q=10^{14}\,\mathrm{GeV}$, we obtain a slightly weaker bound $\kappa_1<0.13$.

We turn now to the other case, $m_L^2<m_E^2$ .  In a small parameter range it also gives stau as the NLSP, but we shall not discuss this possibility in detail; rather,  this case is mainly interesting because it can give  sneutrino NLSP. The $\tau$ sneutrino is the lightest due to the  $\tau$ Yukawa coupling driving its mass slightly below the others.  The following three conditions are  relevant for constraining the parameter space:  $m_L^2< \{m_E^2, M_1^2, M_2^2\}$. Using  Appendix A we get the bounds $\kappa_1<0.24-1.8\kappa_2$  and $\kappa_2<0.09$, implying, in particular, $\kappa_1<0.24$, and that the bino and wino physical masses must be in the range $0.6M_1<M_2<1.2M_1$. For sufficiently large left-right mass splitting compared to the left-right mass mixing term of the slepton mass matrix we get a sneutrino NLSP. This is enabled by  the electroweak $D$-term contributions  to slepton and sneutrino masses. For moderate $\tan\beta$ the sneutrino mass after EWSB reads
\begin{equation}
m^2_{\tilde \nu}=m^2_L-\frac{1}{2}M_Z^2\, ,
\end{equation}
whereas the mass matrix of the charged sleptons is
\begin{equation}
\mathbf{m}^2_{\tilde\tau_{L,R}} =\left( \begin{array}{cc} m_L^2+M_W^2-\frac{1}{2}M_Z^2 & -m_\tau\mu\tan\beta \\
-m_\tau\mu\tan\beta & m_E^2+M_Z^2-M_W^2  \end{array}\right) \, .
\end{equation} 
The $D$-term contributions to sneutrino masses are negative, while analogous contributions to masses of charged sleptons are positive. Hence, one can achieve sneutrino NLSP when $m_L^2<m_E^2$, provided the left-right mixing in the slepton sector does not give too large negative contribution to the lightest charged slepton mass from eigenvalue level repulsion:
For $m_E^2-m_L^2\approx\mathcal{O}(1)m_L^2$ one finds
\begin{equation}
m^2_{\tilde\tau_1} \approx m_L^2 + M_W^2-\frac{1}{2}M_Z^2 -\frac{m_\tau^2\mu^2\tan^2\beta}{m_E^2-m_L^2},
\end{equation}
where $m_{\tilde\tau_1}$ is the lightest eigenvalue of the $\mathbf{m}^2_{\tilde\tau_{L,R}}$ matrix and is mostly $m^2_{\tilde\tau_L}$.
The condition $m_{\tilde\nu}<m_{\tilde\tau_1}$ can be expressed as
\begin{equation}
m_E^2-m_L^2 > \frac{m_\tau^2\mu^2\tan^2\beta}{m_W^2} \, .
\end{equation}
With $\mu=1000\,\mathrm{GeV}$ and $\tan\beta=10$ this requires a minimal splitting between $\sqrt{m_E^2}$ and $\sqrt{m_L^2}$ of about $100\,\mathrm{GeV}$.

For both stau or sneutrino as the NLSP, further contraints on the parameter space arise. In 
the MSSM, with small $A$-terms, the lightest Higgs boson mass generally needs to be above the $m_h>114\gev$ experiment limit. This is accomplished by setting a lower bound of about 1 TeV on the geometrical average of the stop masses. Another important constraint is requiring that the higgsino be heavier than the NLSP.  Both constraints put some bounds on the parameter $\kappa_3$ and their relative importance depends of the ratios of various mass scales.

It is well know that the main renormalization  effect on the stop masses comes from the gluino contribution. Thus,  heavy stops in the presence of light gluinos imply  large initial values of the stop masses (large $\kappa_3$). Let $\zeta$ be the ratio of the minimal geometric mean mass of the stops, necessary to satisfy the Higgs mass bound, to the mass of the degenerate gauginos. Then $\zeta$ can be approximately expressed as:
\begin{eqnarray}
\label{hico}
\zeta^4 = (1.1-0.27\kappa_1+1.6\kappa_2+0.41\kappa_3)(0.49+1.4\kappa_1+0.28\kappa_3-0.95\kappa_2) \, .
\end{eqnarray}
Solving this quadratic equation for $\kappa_3$, one obtains the values shown in Table \ref{tkap3}.
For instance, for gaugino masses  at 500 GeV   one needs $\kappa_3$ of order 10, independently of the NLSP mass.

\begin{table}
\begin{center}
\begin{tabular}{c|ccccc}
\hline
\hline
& $\kappa_2=0$ & $\kappa_2=0.5$ & $\kappa_2=1$ & $\kappa_2=2$ & $\kappa_2=5$ \\
\hline
$\zeta=1$ & $0.5$ & $1$ & $2$ & $5$ & $15$ \\
$\zeta=1.5$ & $4$ & $4$ & $5$ & $7$ & $16$ \\
$\zeta=2$ & $9$ & $9$ & $10$ & 11 & 18 \\
$\zeta=3$ & 24 & 24 & 24 & 25 & 29 \\
\hline
\hline
\end{tabular}
\end{center}
\caption{\em  We show the values of $\kappa_3$ satisfying the Higgs boson mass constraint  parametrized by eq.~(\ref{hico}). The results are relatively insensitive to $\kappa_1$, but the value of  $\kappa_1=0.08$ was chosen to construct this table. \label{tkap3}} 
\end{table}

Another important constraint is that the NLSP  is lighter than the higgsino. The two are related to each other by the condition of proper electroweak breaking and, with a light gluino in the spectrum  one 
obtains a constraint on the soft stop and Higgs masses. In the minimum of the Higgs potential one has approximately (among other approximations, we neglect the running of $\mu$)
\begin{equation}
\label{c6}
\mu^2\approx  -m^2_{H_2}.
\end{equation}
If $m_\mathrm{NLSP}=a M_1$ and the gaugino spectrum is degenerate, $M_1=M_2=M_3$, 
the requirement that the higgsino is heavier than the NLSP can be approximately expressed as
\begin{equation}
\kappa_3 > 1.73\kappa_2+2.7a^2 .
\end{equation}

\begin{figure}
\begin{center}
\includegraphics*[height=6.6cm]{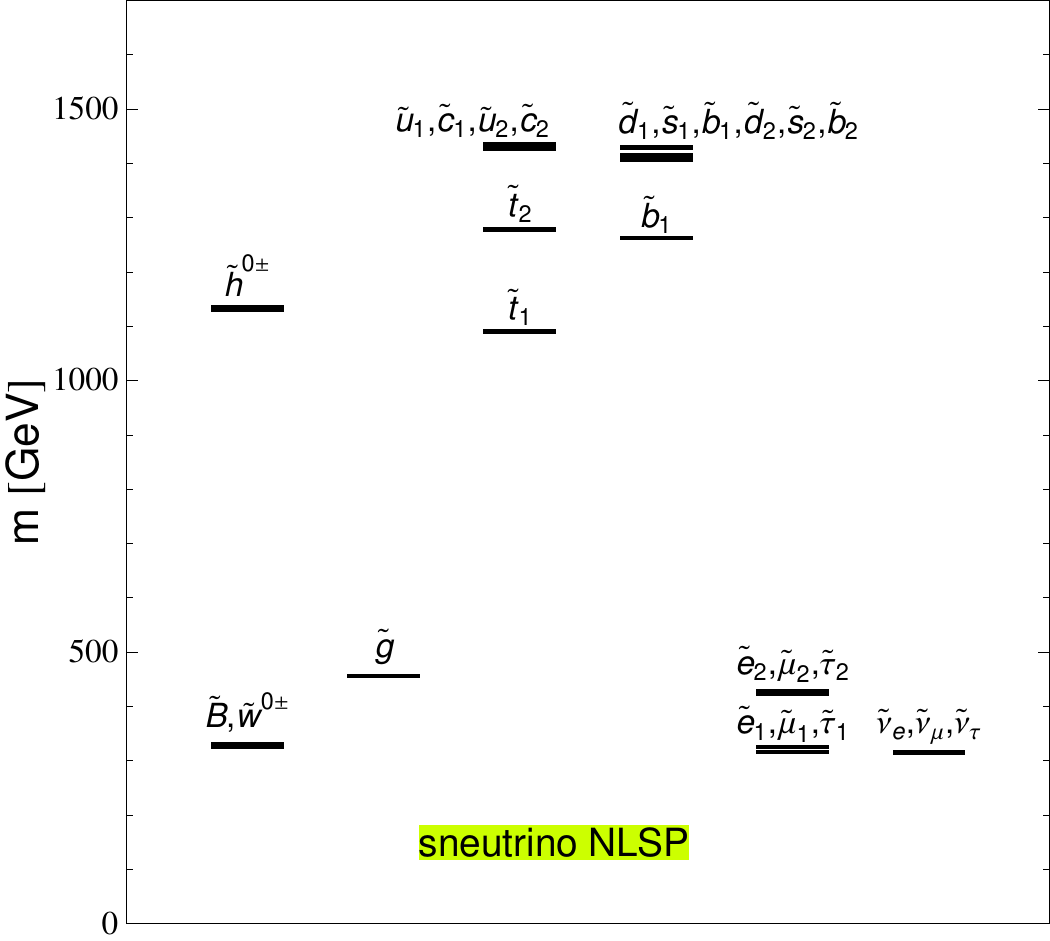}
\hspace{0.5cm}
\includegraphics*[height=6.6cm]{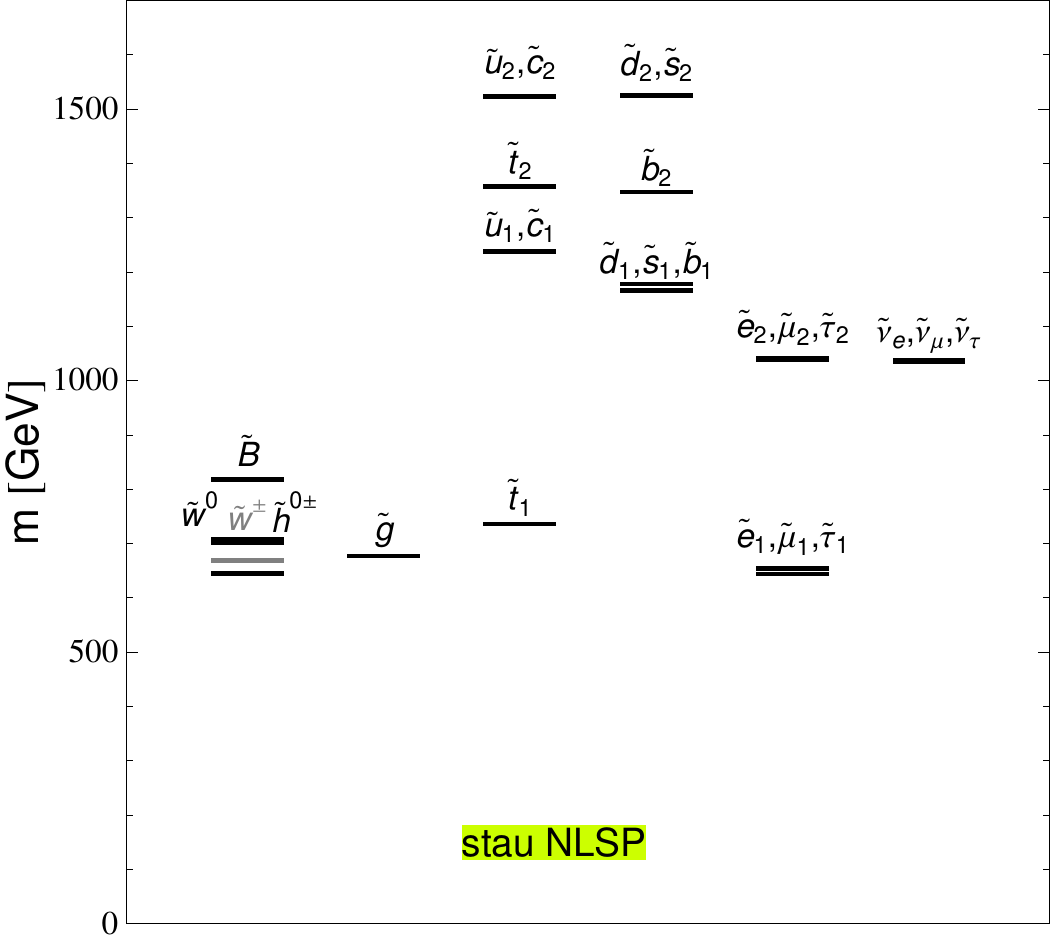}
\end{center}
\caption{\em  The MSSM spectra of models given by the parameter sets shown in Table \ref{tsp} for
the sneutrino NLSP (left) and the stau NLSP (right). \label{fsp}}
\end{figure}

Similarly as for the other constraint, the larger  the electroweak doublet contribution to the scalar masses the larger the coloured contribution must be. Furthermore, since as we have shown earlier, the parameter $a$ must be not far from 1, the higgsino constraint gives us $\kappa_3$ at least of about 3, independently of all the mass scales. In summary, in model building, one has to arrange for large contributions to the soft masses of the coloured particles (large $\kappa_3$)  while suppressing the doublet and hypercharge contributions (small $\kappa_2$ and $\kappa_1$). Examples of the parameter sets giving the sneutrino and stau as the NLSP are given in Table \ref{tsp} and the corresponding mass spectra calculated with the {\tt suspect.2.3} code \cite{Djouadi:2002ze} are shown in Fig.~\ref{fsp}. The result is a rather compressed MSSM superpartner spectrum, which bears some resemblance to the spectra of other forms of  `compressed supersymmetry' studied in different contexts~\cite{Kane:1998im,BasteroGil:1999gu,Martin:2007gf}.  We emphasize that obtaining these examples  is not difficult, and the parameter sets of Table~\ref{tsp} are not special. There are many possible spectra that satisfy the constraints subject to the general qualitative pattern we have discussed. Our specific examples serve mainly to give concrete spectra for illustration purposes later.

\begin{table}
\begin{center}
\begin{tabular}{c|cccc|cc}
\hline
\hline
& $\Lambda_3/\Lambda_1$ & $\Lambda_2/\Lambda_1$ & $\Lambda_1$ & $(\kappa_1,\kappa_2,\kappa_3)$ & $(M_1,M_2,M_3)$ & $m_\mathrm{NLSP}$ \\
\hline
sneutrino NLSP & 0.2 & 0.52 & 240 TeV & (0.22,0.042,28) & (325,330,456) GeV & 314 GeV \\
stau NLSP & 0.16 & 0.6 & 500 TeV & (0.07,0.5,6) & (818,645,677) GeV & 644 GeV \\
\hline
\hline
\end{tabular}
\end{center}
\caption{\em Two exemplary parameter sets giving sneutrino and stau NLSP. 
\label{tsp}}
\end{table}

\section{Messenger sector of the general gauge mediation models}

We can now ask, which choices of the messenger sector in general gauge mediation models
can lead to the pattern of the soft masses characterized in the preceding Section. 
We begin by outlining the general procedure and later illustrate it with the special case of the stau NLSP of Table~\ref{tsp}. Let us consider a general gauge mediation model with the messenger sector consisting of $N$ messenger pairs\footnote{We define ``messenger pair" to mean $R+\bar R$ for complex representations and just $R$ for self-conjugate representations. When we write $R+\bar R$ below, one should read it as just $R$ if $R$ is self-conjugate.}  $R^{(a)}+\bar R^{(a)}$, where $R^{(a)}$, $a=1,\ldots,N$, is a representation of
$SU(5)$. $R^{(a)}$ can be decomposed into $N_a$ irreducible Standard Model representations $R^{(a)}_1,\ldots,R^{(a)}_{N_a}$.
The influence of each Standard Model representation to the soft masses can be characterized by
a mass scale $\xi_{i,a}$, where $i=1,\ldots,N_a$ and $a=1,\ldots,N$. The expressions for $\Lambda_r$ and $\tilde\Lambda_r^2$, parametrizing general gauge mediation, are 
\begin{eqnarray}
\Lambda_3 &=& \sum_{a=1}^N \sum_{i=1}^{N_a} d_{i,a} \xi_{i,a}\,,\qquad
\Lambda_2 = \sum_{a=1}^N \sum_{i=1}^{N_a} d'_{i,a} \xi_{i,a} \,,\qquad
\Lambda_1 = \sum_{a=1}^N \sum_{i=1}^{N_a} Y^2_{i,a} \xi_{i,a} \\
\tilde\Lambda_3^2 &=& \sum_{a=1}^N \sum_{i=1}^{N_a} d_{i,a}\xi_{i,a}^2 \,,\qquad
\tilde\Lambda_2^2 = \sum_{a=1}^N \sum_{i=1}^{N_a} d'_{i,a} \xi_{i,a}^2 \,,\qquad
\tilde\Lambda_1^2 = \sum_{a=1}^N \sum_{i=1}^{N_a} Y^2_{i,a} \xi_{i,a}^2 \, .
\end{eqnarray}
Here $d_{i,a}$, $d'_{i,a}$ and $Y_{i,a}^2$ are the Dynkin indices of $R^{(a)}_i$ with respect to $SU(3)$, $SU(2)$ and $U(1)$,
respectively. From now on, we shall mainly discuss examples of just one messenger pair, so we 
drop the index $a$ from $R^{(a)}$ to make our formulae clearer. 
Before discussing specific examples, we start by three simple general observations. 

First, minimizing $\kappa_r$ is equivalent to minimizing $\tilde\Lambda_r^2$ subject to
the constraint $\Lambda_r=\mathrm{constant}$. The minimal value of $\kappa_r$  is obtained  for all $\xi_{i,a}$ equal. That minimal value is $1/D$, where $D$ is the total sum of Dyndin indices of the messenger sector. Such a `universal' solution would provide an absolute lower bound for $\kappa_1$ and it will later be convenient to study properties of the messenger sector that allow for sizable deviations from this solution.

Second, above the scale of the messenger representation $R_i+\overline{R}_i$ 
with Dynkin indices $d_i$, $d'_i$ and $Y_i^2$,
the fields in this representation give extra contributions to the running of the gauge couplings: $\Delta\beta_3=d_ig_3^3/(16\pi^2)$, $\Delta\beta_2=d'_ig_2^3/(16\pi^2)$ and $\Delta\beta_1=Y^2_ig_1^3/(16\pi^2)$. 
The total Dynkin index $D$ of a representation $R+\overline{R}$  is the sum of the Dynkin indices over all component representations for any Standard Model gauge group: 
\begin{equation}
D=\sum_{i} d_i = \sum_{i}d'_i = \sum_{i}Y^2_i\, .
\end{equation}
As can be seen from Tables \ref{t1} and \ref{t2} in Appendix B,
these contributions to the $\beta$-functions can be large and the gauge couplings can enter a nonperturbative regime or even encounter a Landau pole. Since the $SU(3)$ gauge coupling is the largest, we check that for a total messenger Dynkin index larger than 24 or 50, the gauge coupling goes strong (i.e., $g>\sqrt{4\pi}$) at $Q=2\cdot 10^{16}\,\mathrm{GeV}$ for the messenger scale of $10^{14}\,\mathrm{GeV}$ or  $10^{15}\,\mathrm{GeV}$, respectively. We see that a requirement of perturbativity up to the GUT scale
gives us a finite number of possibilities for the particle content of the messenger sector. 

Finally, even if the messenger spectrum consists of complete $SU(5)$ multiplets,
mass splittings within the multiplets may affect the running of gauge couplings,
potentially spoiling unification.  With the MSSM gauge couplings unifying to a good accuracy in the
absence of the messenger thresholds, one can account for how much the messengers
spoil unification, 
\begin{equation}
\label{etadef}
\eta_{rr'}(\Lambda_U)\equiv \frac{1}{\alpha_r(\Lambda_U)}-\frac{1}{\alpha_{r'}(\Lambda_U)}=\sum_i \frac{(b_{ir}-b_{ir'})}{2\pi}\ln \frac{\Lambda_U}{\mathcal{M}_i} \, ,
\end{equation}
where the index $r,r'$ denotes the gauge group, i.e.\ $r,r'=1,2,3$ is for $U(1)$, $SU(2)$ and $SU(3)$. The index $i$ runs over Standard Model representations,  $\mathcal{M}_i$ is the mass of the $i^{\rm th}$ representation, and $b_{ir}$ is the contribution to the $\beta$ function of $\alpha_r$ from the $i^{\rm th}$ representation, where
$\mathrm{d}\alpha/\mathrm{d}\ln Q=(b/(2\pi))\alpha^2$
sets the normalization of $b$. $\Lambda_U$ can be any reasonably defined unification scale, such as the scale at which $\alpha_1(\Lambda_U)=\alpha_2(\Lambda_U)$.
Note that for complete messenger representations $\sum_i b_{ir}$ does not depend on $i$ and
$\eta_{rr'}$ is then a function of the masses and Dynkin indices of the fields in the messenger sector, but not of the unification scale $\Lambda_U$.

We are now ready to discuss a few specific examples. Finding a messenger spectrum of any specific model, whether with stau NLSP or sneutino NLSP, can be accomplished by the general techniques described above. For the sake of concreteness we choose the stau NLSP model of table~\ref{tsp} to illustrate the general procedure.  We shall concentrate on finding the `simplest' or `most minimal' messenger model for this spectrum. The  identification of what is minimal can be approached in several ways. Here we define three different characterizations: (i) the smallest dimension of the non self-conjugate representations used, (ii) the smallest Dynkin index, and (iii) the smallest number of Standard Model representations employed, i.e.\ the smallest number of free parameters.
These three models are summarized in Table \ref{tmess}.

\begin{table}
\begin{center}
\begin{tabular}{c|cc|cc}
\hline
\hline
model & \# of copies & $SU(5)$ representation & total Dynkin index & \# of parameters \\  
\hline
(i) & 1 & $\mathbf{40}+\mathbf{\overline{40}}$ & 22 & 6 \\
(ii) & 3 & $\mathbf{24}$ & 15 & 9 \\
(iii) & 1 & $\mathbf{75}$ & 25 & 5 \\
\hline
\hline
\end{tabular}
\end{center}
\caption{\em Three different choices for a minimal messenger sector that can produce the required gauge-mediated spectrum derived in sec.~\ref{UV Conditions}. \label{tmess}}
\end{table}

Model (i). With one pair of $\mathbf{40}+\mathbf{\overline{40}}$ we obtain solutions corresponding to
the stau NLSP parameter set in Table \ref{tsp}. Two example sets of $\xi_{i,a}$ values are collected in Table \ref{t40}.
These features follow from the following properties of the $\mathbf{40}+\mathbf{\overline{40}}$
representation. In order to keep the value of $\kappa_1$ close to the absolute minimal value of
$\sim 0.045$, the representations with large $Y_i^2$, i.e.\ $(\mathbf{1},\mathbf{2},-9)$, $(\mathbf{\bar3},\mathbf{3},-4)$ and $(\mathbf{8},\mathbf{1},6)$, should not deviate too much from the `universal' solution of $\xi_{i,a}={\rm const}$. The existence of representations with small $Y^2_i$ and large $d_i$ is crucial for obtaining the spectra with light gluinos and heavy squarks. It is this reason, for example, that a model employing one pair of $\mathbf{45}+\mathbf{\overline{45}}$ does not have solutions for a stau NLSP
parameter set from Table \ref{tsp}.
The prospects for exact gauge coupling unification can be directly checked for each set of solutions. 
The computed values of $\eta_{rr'}$ for each of the given solutions are presented in Table~\ref{t40}.

\begin{table}
\begin{center}
\begin{tabular}{c|cccccc|cc}
\hline
\hline
solution &
$\xi_{(\mathbf{1},\mathbf{2},-9)}$ &
$\xi_{(\mathbf{3},\mathbf{2},1)}$ &
$\xi_{(\mathbf{\bar 3},\mathbf{1},-4)}$ &
$\xi_{(\mathbf{\bar 3},\mathbf{3},-4)}$ &
$\xi_{(\mathbf{8},\mathbf{1},6)}$ &
$\xi_{(\mathbf{\bar 6},\mathbf{2},1)}$ &
$\eta_{12}$ & $\eta_{13}$ \\
\hline
1 & 0.077 & 0.22 & 0.092 & 0.020 & 0.033& $-0.063$ &
$-0.015$ & 0.85\\
2 & 0.077 & 0.22 & $-0.011$ & 0.020 & 0.050 & $-0.063$ &
$-0.13$ & 0.80 \\
\hline
\hline
\end{tabular}
\end{center}
\caption{\em $\xi_{i,a}$ solutions for $\mathbf{40}+\mathbf{\overline{40}}$ in units of $\Lambda_1$ for the stau NLSP parameter set from Table \ref{tsp}.  \label{t40}}
\end{table}

Model (ii).
With three copies of $\mathbf{24}$, we have 9 free parameters. We discard one parameter by assuming
that $\xi_{(\mathbf{3},\mathbf{2},5),1}=\xi_{(\mathbf{3},\mathbf{2},5),2}$ and then
fix all $\xi_{(\mathbf{3},\mathbf{2},5),a}$, $a=1,2,3$, by solving for $\kappa_1$ for fixed $\Lambda_1$. The minimal possible value of $\kappa_1$ is $1/15\approx0.067$, close to the
value used for the stau NLSP parameter set from Table \ref{tsp}.  Two out of
three $\xi_{(\mathbf{8},\mathbf{1},0),a}$ ($\xi_{(\mathbf{1},\mathbf{3},0),a}$) can be used for adjusting $\Lambda_3$ and $\kappa_3$ ($\Lambda_2$ and $\kappa_2$), and the two remaining parameters can be set at values giving desired $\eta_{rr'}$. Solutions with $\eta_{rr'}=0$ are shown in
Table \ref{t24}.

We see that the messenger sector consisting of copies of $\mathbf{24}$ is very flexible, i.e.\ one can
obtain a wide class of the GGM parameter sets, but at the same time rather not predictive. 
This is because $\mathbf{24}$ contains sets of fields charged only with respect to $SU(3)$ and
$SU(2)$, with vanishing hypercharge. This freedom 
can be utilized to build more ambitious models. Since an adjoint Higgs field $\mathbf{24}$ is used to
break the unified gauge symmetry in minimal $SU(5)$ GUTs, one can, in principle,
achieve GUT symmetry breaking and supersymmetry breaking within the same sector.
A simple model build along these lines, with three copies of $\mathbf{24}$, can be found in \cite{Bajc:2008vk}, though its MSSM spectrum is very different from what we assume here, as the 
model in~\cite{Bajc:2008vk} predicts vanishing gaugino masses at one loop.

\begin{table}
\begin{center}
\begin{tabular}{c|cccccccc}
\hline
\hline
solution &
$\xi_{(\mathbf{3},\mathbf{2},-5),1,2}$ &
$\xi_{(\mathbf{3},\mathbf{2},-5),3}$ &
$\xi_{(\mathbf{1},\mathbf{3},0),1}$ &
$\xi_{(\mathbf{1},\mathbf{3},0),2}$ &
$\xi_{(\mathbf{1},\mathbf{3},0),3}$ &
$\xi_{(\mathbf{8},\mathbf{1},0),1}$ &
$\xi_{(\mathbf{8},\mathbf{1},0),2}$ &
$\xi_{(\mathbf{8},\mathbf{1},0),3}$ \\
\hline
1 & 0.056 & 0.088 &  0.18 & $ -0.19$ & 0.0080 & 0.087 & $-0.18$ & 0.012 \\
2 & 0.077 & 0.046 & 0.18 & $-0.19$ & 0.0080 & 0.087 & $-0.18 $ & 0.012 \\
\hline
\hline
\end{tabular}
\end{center}
\caption{\em  $\xi_{i,a}$ solutions for three copies of $\mathbf{24}$ in units of $\Lambda_1$ for the stau NLSP parameter set from Table \ref{tsp}. These solutions yield $\eta_{rr'}=0$.  \label{t24}}
\end{table}

Model (iii). 
With one $\mathbf{75}$ we have only 5 parameters, and two solutions are presented in Table \ref{t75}. With this messenger sector it is straightforward to get $\kappa_2$ ranging in value from 0.052 to 0.16, covering what is needed to reproduce Table \ref{tsp}. 
The smallness of $\Lambda_3$ is achieved mainly thanks to a large negative value of $\xi_{(\mathbf{8},\mathbf{1},0)}$ which partially
cancels the contribution of $\xi_{(\mathbf{8},\mathbf{3},0)}$ and $\xi_{(\mathbf{6},\mathbf{2},5)}$, and the latter, having the largest $d'_i$, are constrained by the required values of $\Lambda_2$. There are also solutions for $\Lambda_3,\Lambda_2<0$, for which a negative
contribution from $\xi_{(\mathbf{8},\mathbf{3},0)}$ to $\Lambda_2$ dominates over negative $\xi_{(\mathbf{8},\mathbf{1},0)}$ and partially cancels the contribution from $\xi_{(\mathbf{6},\mathbf{2},5)}$. Similarly as in model (ii), desired parameters can be obtained easily, since there are
two  representations with zero hypercharge in $\mathbf{75}$, though in this case we are more restricted by
the $SU(2)$ sector, since all representations charged under $SU(2)$ also carry $SU(3)$ charge.

\begin{table}
\begin{center}
\begin{tabular}{c|ccccc|c|cc}
\hline
\hline
solution &
$\xi_{(\mathbf{8},\mathbf{3},0)}$ &
$\xi_{(\mathbf{8},\mathbf{1},0)}$ &
$\xi_{(\mathbf{3},\mathbf{1},10)}$ &
$\xi_{(\mathbf{3},\mathbf{2},-5)}$ &
$\xi_{(\mathbf{6},\mathbf{2},5)}$ &
$\kappa_2$ &  $\eta_{12}$ & $\eta_{13}$ \\
\hline
1 & 0.023 & $-0.20$  & 0.061 &  $-0.029$ & 0.055 & 0.078 &
$-2.2$ & $-0.48$ \\
2 & 0.00020 & $-0.19$ & 0.00054 & 0.090 & 0.055 & 0.12 &
$-7.1$ & $-1.1$ \\
\hline
\hline
\end{tabular}
\end{center}
\caption{\em  $\xi_{i,a}$ solutions for $\mathbf{75}$ in units of $\Lambda_1$ for the stau NLSP parameter set from Table \ref{tsp}.  \label{t75}}
\end{table}

The example models presented above illustrate some general conditions that exist for any solution. First, the total Dynkin index of the messenger sector needs to be large, at least 15. Second, the number of free parameters is more than a few when considering each Standard Model component representation separately. And third, there must exist Standard Model component representations in the solution that have large QCD charge but relatively small hypercharge in order to lift the stop masses without lifting the slepton mass.  The model with three $\mathbf{24}$'s and the model with one $\mathbf{75}$ are especially good in this regard since adjoints of QCD are present that carry no other Standard Model charge.

The messenger model construction carried out in the latter part of this section has been for the stau NLSP example. We remark that models giving sneutrino as the NLSP differ from those described above mainly by the fact they require a small $\kappa_2$. In this case, the most promising choices for the messenger sectors are the  $SU(5)$ representations that contain many SM representations with $SU(2)$ singlets, e.g.\ $\mathbf{45}+\mathbf{\overline{45}}$ or $\mathbf{50}+\mathbf{\overline{50}}$. Those choices provide the messenger spectra needed to minimize $\kappa_2$ while retaining the freedom to adjust other GGM parameters.

And finally, it is in principle possible to construct the messenger sector so that not only
the desired GGM parameters are obtained but also the metastable supersymmetry breaking minimum arises due to interactions between the spurion and the messengers. It is an open question whether our gauge-mediation messenger models are compatible with this approach that has been successful in other contexts~\cite{Cheung:2007es,Lalak:2008bc,Buican:2008ws}.

\section{Discussion and Conclusions}

Our goal has been to find the particle spectrum that allows the maximum reheat temperature so as to give thermal leptogenesis a chance to generate the baryon asymmetry of the universe. In this journey we have had to make choices that accomplish this task while keeping the superpartner spectrum natural, satisfying BBN constraints, explaining cold dark matter, lifting the lightest Higgs mass above the experimental limit, etc. 

The first choice was to assume the gravitino is the LSP. The relic abundance of the gravitino depends on the reheat temperature post inflation and the details of the supersymmetric particle spectrum.  The particle spectrum choice also must then be compared with BBN constraints. Maximizing the reheat temperature whilst remaining in line with BBN constraints resulted in considering stau or sneutrino NLSP and a degenerate spectrum of gaugino masses close to the NLSP mass. Furthermore, the additional constraint of Higgs boson mass limit implied that the top squarks need to be above about 1 TeV. The combination of all these effects led us to  well-defined characteristics of the low-scale spectrum.

The next task was to find gauge mediation models that predict a spectrum with the above characteristics. We concluded that minimal gauge mediation with $\mathbf{5}+\mathbf{\bar 5}$ messenger sector does not work, partly because the gauginos are not degenerate at the low-scale in that model, and also because the top squark to NLSP mass ratio is not high enough. We were led to approach the problem from the viewpoint of generalized gauge-mediation model building.  In the last section we showed how a traditional messenger model can account for the required spectrum, albeit for a spectrum of messenger states that are not commonly employed: $\mathbf{40}+\mathbf{\overline{40}}$, $3\times \mathbf{24}$, or $\mathbf{75}$.

We would like to make some comments on direct collider experimental constraints on the superpartners within these models. In the cases considered, the lifetime of the NLSP is well above the flight time of the NLSP within the detectors. Thus, the NLSP should be considered a stable particle from the point of view of collider physics.  

In the case of sneutrino NLSP, the phenomenology is similar in some ways to standard supersymmetry phenomenology with neutralino LSP that escapes detection. This scenario has been studied at colliders in ref.~\cite{Covi:2007xj}. One important difference with respect to standard supersymmetry phenomenology is the preponderance of leptons and neutrinos accompanying  supersymmetric events that originate even from squark or gaugino production.  A prolific source of leptons and neutrinos in the cascade decays arise from $\tilde l_L\to \tilde\nu_L l\bar \nu$ followed by $\tilde\nu\to \nu\tilde G$. Since $m_{\tilde l_L}$ is close in mass to the $m_{\tilde \nu_L}$ due to being in the same $SU(2)$ multiplet, the leptons coming off these cascade decays are typically rather soft.  Nevertheless, they can be useful not necessarily at the trigger level, where higher momentum leptons are required, but at the analysis level where the identification of extra leptons in a final state reduces backgrounds. 

Recognizing the correlations between neutral and charged left-handed sleptons, the direct limit at LEP2 of sneutrino mass should be somewhat close to the kinematic limit of $\sim 100\gev$.   Direct limits from Tevatron are not better. However,  the most important constraints do not come from the slepton direct production derived limits but from gluino production, since our parameter space prefers degenerate gauginos near the NLSP mass. Current bounds on the gluino at the Tevatron in standard neutralino LSP supersymmetry is $m_{\tilde g}>308\gev$ at the 95\% CL for any squark mass~\cite{Abazov:2007ww} -- a bound that should be close to that of our case as well. Since we require in our parameter space that the gluino be close in mass to the NLSP, this puts a constraint on the NLSP mass. For example, if we insist the gluino be less than twice the NLSP mass, this puts a limit on the NLSP mass of about $m_{\rm NLSP}>150\gev$.
 
 In the case of stau NLSP, the phenomenology is radically different. In this case we have a charged stable particle that has unique signatures within a detector. The direct limit on a charged stau from LEP2 is $98\gev$ at the 95\% CL~\cite{Abbiendi:2003yd,LEPSUSY}. However, again, in our scenario it is not the direct production of the NLSP  that is most important for this theory. What is most important is gluino pair production at the Tevatron, since the gluino is expected to have mass near that of the stau. In that case the gluino production cross-section would be the dominant superpartner rate, and since all superpartners will decay first to the NLSP before subsequently decaying with long lifetime to the gravitino, all superpartner cross-sections, including the gluinos, must be taken into consideration as a potential copious source of staus. For stable stau track searches at the Tevatron, it is estimated by ref.~\cite{Carpenter:2008he}, from results in~\cite{CDFNote8701,Abazov:2008qu} (see also~\cite{Aaltonen:2009ke}),  that the total superpartner production rate at the Tevatron must be less than $10\, {\rm fb}$ in order to escape having been seen. This number is conservative. Using it implies the gluino mass must be greater than about $500\gev$.  So, for example, if we insist that the gluino be less than twice the NLSP mass, this puts a limit on the stau NLSP mass of about $m_{\tilde\tau}>250\gev$ from Tevatron data.
 
The LHC running above $10\tev$ will make a tremendous qualitative improvement on the discovery capabilities of the models we have presented in this paper. In the case of sneutrino LSP, the searches will be for traditional missing energy signatures and with several ${\rm fb}^{-1}$ they will extend the reach into the multi-TeV region of strongly interacting superpartners. In the case of stau LSP, the searches will be even more powerful. Only a few verifiable events of a slowly moving charged stable track in the detector are enough for discovery~\cite{Rizzi:2008}. This promises discovery or strong bounds into the many-TeV region after only a few inverse femtobarns of data at the LHC.

\section*{Acknowledgments}
\vspace*{.5cm}
\noindent This work was partially supported by TOK Project  
MTKD-CT-2005-029466 and the U.S. Department of Energy.
KT is partially supported by the Foundation for Polish Science through its programme Homing.  We gratefully acknowledge stimulating discussions with Laura Covi and Frank Steffen. MO would like to thank MPI, Munich for their hospitality, and SP and KT would like to thank CERN for their hospitality. 
\section*{Appendix A: Hybrid parametrization of the MSSM RGE solutions}

Here we solve the RGE of the MSSM using the approximate 
semi-analytical method put forward in \cite{bu}.
For our purposes, it is  convenient   to use a hybrid parametrization of these solutions,  with  the physical gaugino masses $M_r$ and with $\kappa_i$ defined at the messenger scale. 
For $Q=10^{15}\,\mathrm{GeV}$, the physical gaugino masses are related to the gaugino masses
at the messenger scale $Q$ by 
\begin{equation}
\tilde M_1 = 2.0M_1\,,\qquad \tilde M_2 = 1.2M_2\,,\qquad \tilde M_3 = 0.44M_3\,
\end{equation}
(for $Q=10^{14}\,\mathrm{GeV}$ the coefficients change to 1.9, 1.2, 0.47, respectively). The gaugino-induced masses of scalars can be written as
\begin{equation}
\label{scasol}
m_s^2 =  c_{ij} M_iM_j \, ,
\end{equation}
where the coefficients $c_{ij}$ multiplying $M_iM_j$ in this expression are given in Table
\ref{trge}.

\begin{table}
\begin{center}
\begin{tabular}{c|cccccc}
\hline
\hline
$s$ & $M_1^2$ & $M_2^2$ & $M_3^2$ & $M_1M_2$ & $M_1M_3$ & $M_2M_3$ \\ 
\hline
\hline
\multicolumn{7}{c}{$Q=10^{15}\,\mathrm{GeV}$}\\
\hline
$Q$ & $-0.023-0.27\kappa_1$ & $0.47+1.6\kappa_2$ & $0.62+0.41\kappa_3$ & $-0.0036$ & $-0.0057$ & $-0.022$ \\
$U$ & $0.17+1.4\kappa_1$ & $-0.20-0.95\kappa_2$ & $0.52+0.28\kappa_3$ & $-0.0072$ & $-0.011$ & $-0.044$\\
$D$ & $0.063+0.55\kappa_1$ & 0 & $0.71+0.53\kappa_3$ & 0 & 0 & 0 \\
$L$ & $0.14+1.2\kappa_1$ & $0.57+2.1\kappa_2$ & 0 & 0 & 0 & 0 \\
$E$ & $0.56+4.9\kappa_1$ & 0 & 0 & 0 & 0 & 0 \\
$H_2$ & $0.026+0.0014\kappa_1$ & $0.28+0.64\kappa_2$ & $-0.28-0.37\kappa_3$ & $-0.011$ & $-0.017$ & $-0.066$ \\ 
\hline
\hline
\multicolumn{7}{c}{$Q=10^{14}\,\mathrm{GeV}$}\\
\hline
$Q$ & $-0.016-0.22\kappa_1$ & $0.43+1.6\kappa_2$ & $0.61+0.45\kappa_3$ & $-0.0027$ & $-0.0047$ & $-0.019$ \\
$U$ & $0.14+1.2\kappa_1$ & $-0.17-0.90\kappa_2$ & $0.52+0.32\kappa_3$ & $-0.0055$ & $-0.0093$ & $-0.038$\\
$D$ & $0.050+0.46\kappa_1$ & 0 & $0.70+0.58\kappa_3$ & 0 & 0 & 0 \\
$L$ & $0.11+1.0\kappa_1$ & $0.51+2.0\kappa_2$ & 0 & 0 & 0 & 0 \\
$E$ & $0.45+4.2\kappa_1$ & 0 & 0 & 0 & 0 & 0 \\
$H_2$ & $0.026+0.032\kappa_1$ & $0.26+0.66\kappa_2$ & $-0.27-0.39\kappa_3$ & $-0.0082$ & $-0.014$ & $-0.057$ \\ 
\hline
\hline
\end{tabular}
\end{center}
\caption{\em Coefficients of the solution (\ref{scasol}) of the RGE for the scalar masses.  \label{trge}}
\end{table}

\section*{Appendix B: Rudiments of $SU(3)\times SU(2)\times U(1)\subset SU(5)$ group theory}

For convenient reference, in Tables \ref{t1} and \ref{t2} we collect the $(SU(2),SU(3))_{U(1)}$
decomposition of the lowest dimensional representations of $SU(5)$ taken from
\cite{Slansky:1981yr}, listing explicitly the Dynkin indices of each irreducible representation of the
Standard Model gauge group. In Table \ref{t1}, we show the representations of $SU(5)$
which do not contain any zero-hypercharge field. Each of these representations $R$ is accompanied
in the messenger sector by its conjugate partner $\overline{R}$ and the listed Dynkin indices
correspond to the $R+\overline{R}$ pair. In Table \ref{t2}, we show the representations of $SU(5)$
containing at least one field with zero hypercharge. These representations are self-conjugate and the Dynkin indices correspond to only the representations listed in the table.


\begin{table}
\begin{center}
\begin{tabular}{|c|c|ccc|}
\hline
\hline
$SU(5)$ & SM & & & \\
rep. & reps. & $d_i$ & $d'_i$ & $Y_i^2$ \\
\hline
\hline
$\mathbf{5}$ & $(\mathbf{1},\mathbf{2},3)$ & 0 & 1 & 3/5 \\
${}_{D(\mathbf{5}+\mathbf{\overline{5}})=1}$ & $(\mathbf{3},\mathbf{1},-2)$ & 1 & 0 & 2/5\\
\hline
$\mathbf{10}$ & 
$(\mathbf{3},\mathbf{2},1)$ & 2 & 3 & 1/5 \\
${}_{D(\mathbf{10}+\mathbf{\overline{10}})=3}$ &$(\mathbf{\bar 3},\mathbf{1},-4)$ & 1 & 0 & 8/5 \\
&$(\mathbf{1},\mathbf{1},6)$ & 0 & 0 & 6/5 \\
\hline
$\mathbf{15}$ 
& $(\mathbf{1},\mathbf{3},6)$ & 0 & 4 & 18/5 \\
${}_{D(\mathbf{15}+\mathbf{\overline{15}})=7}$
& $(\mathbf{3},\mathbf{2},1)$ & 2 & 3 & 1/5 \\
& $(\mathbf{6},\mathbf{1},-4)$ & 5 & 0 & 16/5 \\
\hline
$\mathbf{35}$
& $(\mathbf{1},\mathbf{4},-9)$ & 0 & 10 & 54/5 \\
${}_{D(\mathbf{35}+\mathbf{\overline{35}})=28}$
& $(\mathbf{\bar 3},\mathbf{3},-4)$ & 3 & 12 & 24/5 \\
& $(\mathbf{\bar 6},\mathbf{2},1)$ & 10 & 6 & 2/5 \\
& $(\mathbf{\overline{10}},\mathbf{1},-6)$ & 15 & 0 & 12 \\
\hline
$\mathbf{40}$
& $(\mathbf{1},\mathbf{2},-9)$ & 0 & 1 & 27/5 \\
${}_{D(\mathbf{40}+\mathbf{\overline{40}})=22}$
& $(\mathbf{3},\mathbf{2},1)$ & 2 & 3 & 1/5 \\
& $(\mathbf{\bar 3},\mathbf{1},-4)$ & 1 & 0 & 8/5 \\
& $(\mathbf{\bar 3},\mathbf{3},-4)$ & 3 & 12 & 24/5\\
& $(\mathbf{8},\mathbf{1},6)$ & 6 & 0 & 48/5 \\
& $(\mathbf{\bar 6},\mathbf{2},1)$ & 10 & 6 & 2/5  \\
\hline
$\mathbf{45}$
& $(\mathbf{1},\mathbf{2},3)$ & 0 & 1 & 3/5 \\
${}_{D(\mathbf{45}+\mathbf{\overline{45}})=24}$
& $(\mathbf{1},\mathbf{1},-2)$ & 1 & 0 & 2/5 \\
& $(\mathbf{3},\mathbf{3},-2)$ & 3 & 12 & 6/5 \\
& $(\mathbf{\bar 3},\mathbf{1},8)$ & 1 & 0 & 32/5 \\
& $(\mathbf{\bar 3},\mathbf{2},-7)$ & 2 & 3 & 49/5 \\
& $(\mathbf{\bar 6},\mathbf{1},-2)$ & 5 & 0 & 4/5 \\
& $(\mathbf{8},\mathbf{2},3)$ & 12 & 8 & 24/5 \\
\hline
\end{tabular}
\begin{tabular}{|c|c|ccc|}
\hline
\hline
$SU(5)$ & SM & & & \\
rep. & reps. & $d_i$ & $d'_i$ & $Y_i^2$ \\
\hline
\hline
$\mathbf{50}$ &
$(\mathbf{1},\mathbf{1},-12)$ & 0 & 0 & 24/5 \\
${}_{D(\mathbf{50}+\mathbf{\overline{50}})=35}$
& $(\mathbf{3},\mathbf{1},-2)$ & 1 & 0 & 2/5 \\
& $(\mathbf{\bar 3},\mathbf{2},-7)$ & 2 & 3 & 49/5\\
& $(\mathbf{\bar 6},\mathbf{3},-2)$ & 15 & 24 & 12/5 \\
& $(\mathbf{6},\mathbf{1},8)$ & 5 & 0 & 64/5 \\
& $(\mathbf{8},\mathbf{2},3)$ & 12 & 8 & 24/5 \\
\hline
$\mathbf{70}$ 
& $(\mathbf{1},\mathbf{2},3)$ & 0 & 1 & 3/5 \\
${}_{D(\mathbf{70}+\mathbf{\overline{70}})=49}$
& $(\mathbf{1},\mathbf{4},3)$ & 0 & 10 & 6/5 \\
& $(\mathbf{3},\mathbf{1},-2)$ & 1 & 0 & 2/5 \\
& $(\mathbf{3},\mathbf{3},-2)$ & 3 & 12 & 6/5 \\
& $(\mathbf{\bar3},\mathbf{3},8)$ & 3 & 12 & 96/5 \\
& $(\mathbf{6},\mathbf{2},-7)$ & 10 & 6 & 98/5 \\
& $(\mathbf{8},\mathbf{2},3)$ & 12 & 8 & 24/5 \\
& $(\mathbf{15},\mathbf{1},-2)$ &20 & 0 & 2 \\
\hline
$\mathbf{70'}$
& $(\mathbf{1},\mathbf{5},-12)$ & 0 & 20 & 24 \\
${}_{D(\mathbf{70'}+\mathbf{\overline{70}'})=84}$
& $(\mathbf{\bar 3},\mathbf{4},-7)$ & 4 & 30 & 98/5 \\
& $(\mathbf{\bar 6},\mathbf{3},-2)$ & 15 & 24 & 12/5 \\
& $(\mathbf{\overline{10}},\mathbf{2},3)$ & 30 & 10 & 6 \\
& $(\mathbf{\overline{15}'},\mathbf{1},8)$ & 35 & 0 & 32 \\
\hline
\multicolumn{5}{c}{} \\
\multicolumn{5}{c}{} \\
\multicolumn{5}{c}{} \\
\multicolumn{5}{c}{} \\
\multicolumn{5}{c}{} \\
\multicolumn{5}{c}{} \\
\end{tabular}
\end{center}
\caption{Lowest dimensional representations of $SU(5)$ for which all component fields are charged under hypercharge. The $SU(3)$ Dynkin indices are $d_i$, the $SU(2)$  are $d'_i$, and the $U(1)_Y$ are $Y_i^2$.  The Dynkin indices are given for the sum of the field \underline{and} its conjugate. The ordering of the SM representations follows \cite{Slansky:1981yr}:  $SU(2)$ and  $SU(3)$ representations followed by hypercharge in the normalization for which the smallest possible hypercharge among all branchings is 1.  \label{t1}}
\end{table}

\begin{table}
\begin{center}
\begin{tabular}{|c|c|ccc|}
\hline
\hline
$SU(5)$ & SM & & & \\
rep. & reps. & $d_i$ & $d'_i$ & $Y_i^2$ \\
\hline
\hline
$\mathbf{24}$ 
& $(\mathbf{1},\mathbf{1},0)$ & 0 & 0 & 0 \\
${}_{D(\mathbf{24})=5}$
& $(\mathbf{1},\mathbf{3},0)$ & 0 & 2 & 0 \\
& $(\mathbf{8},\mathbf{1},0)$ & 3 & 0 & 0 \\
& $(\mathbf{3},\mathbf{2},-5)+\mathrm{c.c.}$  & 2 & 3 & 5  \\
\hline
\multicolumn{5}{c}{} \\
\multicolumn{5}{c}{} \\
\end{tabular}
\begin{tabular}{|c|c|ccc|}
\hline
\hline
$SU(5)$ & SM & & & \\
rep. & reps. & $d_i$ & $d'_i$ & $Y_i^2$ \\
\hline
\hline
$\mathbf{75}$
& $(\mathbf{1},\mathbf{1},0)$ & 0 & 0 & 0 \\
${}_{D(\mathbf{75})=25}$
& $(\mathbf{8},\mathbf{3},0)$ & 9 & 16 & 0 \\
& $(\mathbf{8},\mathbf{1},0)$ & 3 & 0 & 0 \\
& $(\mathbf{3},\mathbf{1},10)+\mathrm{c.c.}$  & 1 & 0 & 10 \\
& $(\mathbf{3},\mathbf{2},-5)+\mathrm{c.c.}$ & 2 & 3 & 5 \\
& $(\mathbf{6},\mathbf{2},5)+\mathrm{c.c.}$ & 10 & 6 & 10 \\
\hline
\end{tabular}
\end{center}
\caption{Lowest dimensional representations of $SU(5)$ for which there exists at least one component representation not charged under hypercharge. The $SU(3)$ Dynkin indices are $d_i$, the $SU(2)$  are $d'_i$, and the $U(1)_Y$ are $Y_i^2$. These representations are self-conjugate, and the Dynkin indices given are for the single field listed. The ordering of the SM representations follows \cite{Slansky:1981yr}:  $SU(2)$ and  $SU(3)$ representations followed by hypercharge in the normalization for which the smallest possible hypercharge among all branchings is 1.\label{t2}}
\end{table}

\end{document}